\documentclass[preprint2]{aastex} 

\shorttitle{X-ray variability of $\sigma$~Orionis young stars with {\em ROSAT}}
\shortauthors{Caballero et al.}

\begin{document}

\title{X-ray variability of $\sigma$~Orionis young stars \\ as observed
with {\em ROSAT}} 

\author{J. A. Caballero\altaffilmark{1}, 
J.~L\'opez-Santiago,
E.~de~Castro and
M.~Cornide}
\affil{Departamento de Astrof\'{\i}sica y Ciencias de la Atm\'osfera, Facultad
de F\'{\i}sica, \\ 
Universidad Complutense de Madrid, E-28040 Madrid, Spain}
\email{caballero@astrax.fis.ucm.es}

\altaffiltext{1}{Investigador Juan de la Cierva at the UCM.}

\begin{abstract}
We used the Aladin Virtual Obsrvatory tool and High Resolution Imager {\em
ROSAT} archival data to search for X-ray variability in scale of days in 23
young stars in the $\sigma$~Orionis cluster and a background galaxy.  
Five stars displayed unambiguous flares and had probabilities $p_{\rm var}
\gg$ 99\,\% of being actual variables.
Two of the detected flares were violent and long-lasting, with maximum
duration of six days and amplitude of eight times above the quiescent level.
We classified another four stars as possible X-ray variables, including the
binary system formed by the B2Vp star $\sigma$~Ori~E and its close late-type
companion.  
This makes a minimum frequency of high-amplitude X-ray variability in
excess of a day of 39\,\% among $\sigma$~Orionis stars.
The incidence of this kind of X-ray variability seems to be lower among
classical T~Tauri stars with mid-infrared flux excesses than among
fast-rotating, disk-less young stars.
\end{abstract}


\section{Introduction}

The \object{$\sigma$~Orionis} cluster ($\tau \sim$ 3\,Ma, $d \sim$ 385\,pc) is
routinely investigated for deriving basic properties associated to the star
formation process: initial mass function, frequency of disks, or incidence of
X-ray emission (Caballero 2008b; Walter et~al. 2009).
In particular, the X-ray emission from early- and late-type cluster stars have
been investigated by a number of authors (Wolk et~al. 1996; Sanz-Forcada et~al.
2004; Franciosini et~al. 2006; Skinner et~al. 2008; L\'opez-Santiago \&
Caballero 2008 and references therein).
However, the X-ray variability in the cluster, which informs on the frequency of
flares, strength of magnetic fields, evolution of angular momenta,
variations in the radiative wind shocks, or presence of a low-mass companion,
has been mostly studied in relatively short time scales, of up 
to 24\,h (with {\em XMM-Newton} and {\em Chandra}; references above).  
Only Bergh\"ofer \& Schmitt (1994) and Groote \& Schmitt (2004) have carried out
mid-term X-ray variability studies in $\sigma$~Orionis (about 75\,ks
distributed over more than 30 days), but just of the two brightest stars: 
$\sigma$~Ori~AF--B (O9.5V + B0.0V + B0.5V) and $\sigma$~Ori~E (B2Vp + ?). 
Here we revisit {\em R\"ontgensatellit} ({\em ROSAT}) data to search for X-ray
variability of $\sigma$~Orionis stars in time scales of days.

\section{Analysis}

\subsection{{\em ROSAT}, HRI, and 1RXH}

{\em ROSAT} (Jun 1990--Feb 1999) was a joint German, US, and UK space telescope
designed for measurement of soft X-rays in the energy range of
0.1--2.4\,keV, corresponding to wavelengths of about 120--6\,\AA.  
One of its focal plane instruments was the {\em ROSAT} High Resolution
Imager (HRI), which provided simultaneosuly relative large field of view
(38$\times$38\,arcmin$^2$) and good spatial resolution
(0.499$\pm$0.001\,arcsec\,pixel$^{-1}$, FWHM $\sim$ 2\,arcsec).
However, HRI had negligible energy resolution.

In this work, we used the {\em ROSAT} Source Catalog of Pointed Observations
with HRI (1RXH; {\em ROSAT} Consortium 2000)\footnote{\tt
http://cdsarc.u-strasbg.fr/viz-bin/Cat?IX/28A }.
This catalog contains a list of sources detected by the Standard Analysis
Software System in reprocessed, public HRI datasets. 
In addition to a set of source and sequence flags provided by the {\em ROSAT}
data centers in Germany, the US, and the UK, each source in the catalog has
associated basic parameters like coordinates, count rate, or signal-to-noise
ratio.
Regrettably, although exposure time is given with an accuracy of 1\,s,
dates of observation start and finish (labels {\tt begDate} and {\tt endDate})
are provided in ``Y:M:D'' mode (date expressed as year, month number, day in
month; e.g. ``1995-02-26'').
As a result, the 1RXH time resolution is of {\em one~day}.
Since we were interested in a qualitative estimation of variability in scales of
days, having a larger precision was unnecessary.

\subsection{Aladin and 1RXH}

\begin{figure*}
\epsscale{2.25}
\plotone{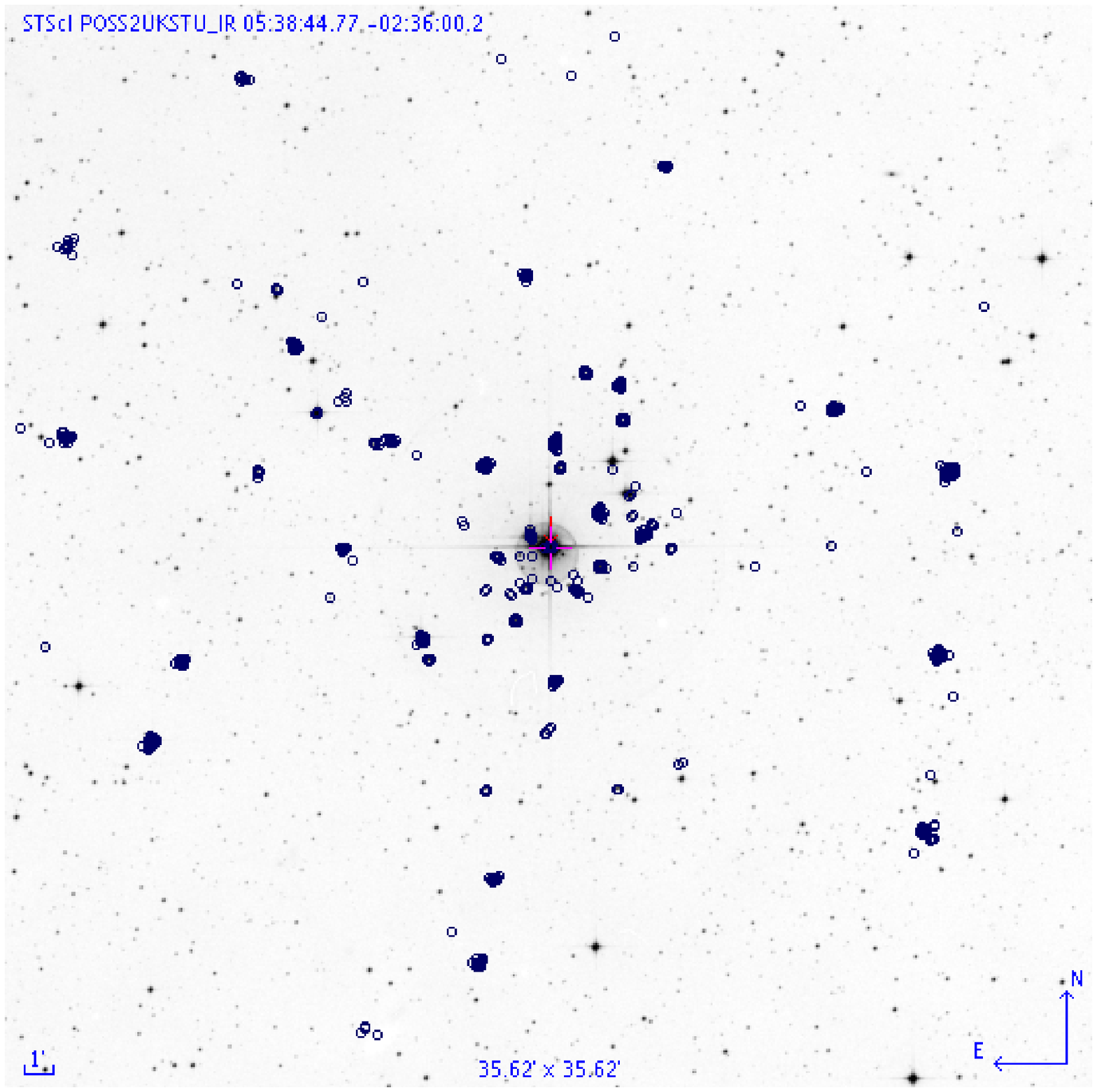}
\caption{1RXH events onto an $I_N$-band digitized image of the Palomar
Observatory Sky Survey-II centered on $\sigma$~Ori~AF--B.
Most of the events, inidicated with small open circles, are associated to
optical counterparts.
Approximate size is $35 \times 35$\,arcmin$^2$, north is up, and east is left.
See the electronic edition of the Journal for a color version of all our
figures. 
\label{fig.HRI}}
\end{figure*}

We loaded all the X-ray 1RXH events at less than 30\,arcmin to
$\sigma$~Ori~AF--B with the Aladin sky atlas (Bonnarel et~al. 2000; 
Caballero 2009).  
For an easier visualization, we overimposed the 1976 detected events onto a
digitized image (Fig.~\ref{fig.HRI}). 

In the 1RXH catalog, there are some single events at angular separations larger
than 30\,arcmin to $\sigma
$~Ori~AF--B, but they are embedded in the nearby
\object{Horsehead Nebula} and \object{Flame Nebula} molecular clouds.
The area investigated by us is, however, nearly free of absorption.
Sherry et~al. (2008) derived $A_V$ = 0.19$\pm$0.02\,mag by comparing the
observed $B-V$ color of a number of stars with the expected $(B-V)_0$ for each
star's spectral type. 
Besides, in the light of some studies of the spatial distribution and
kinematics in the cluster, stars at more than 30\,arcmin to the $\sigma$~Orionis
center have low probabilities to belong to the cluster (Jeffries et~al. 2006;
Caballero 2007a,~2008a). 

\begin{figure}
\epsscale{1.12}
\plotone{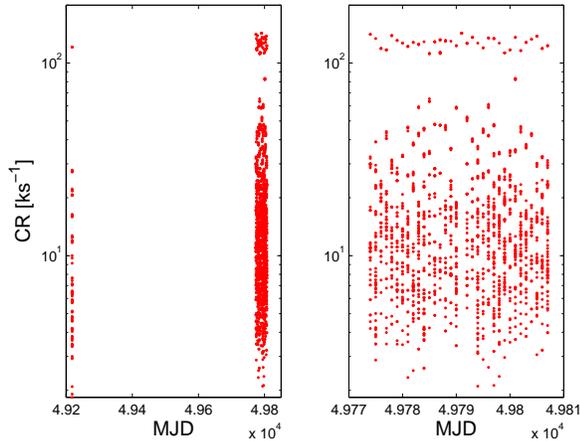}
\caption{Net count error as a function of modified Julian date ($CR$ vs. $t_{\rm
MJD}$).
{\em Left:} The full interval (1992 Sep and 1995 Feb--Mar).  
{\em Right:} Zoom of the 1995 Feb--Mar interval.
Data points above 100\,ks$^{-1}$  correspond to $\sigma$~Ori~AF--B.
\label{fig.crmjd}}
\end{figure}

HRI observations were carried out during two intervals spaced almost three
years. 
The temporal distribution of the observations is illustrated in
Fig.~\ref{fig.crmjd}. 
The main run consisted of one {\em ROSAT} visit per day during 34 days
from 1995~Feb~26 to 1995~Mar~31 (PI: J.~H.~M.~M. Schmitt).
The other run consisted of a single observation on 1992~Sep~13. 
Except for the 1995~Mar~15 observation, which lasted for only 0.22\,ks,
individual ``exposure live times'' during the main run varied in the aprroximate
time interval 1.0--3.6\,ks, with mean and standard deviation {\tt ExpTime} =
2.2$\pm$0.7\,ks.  
However, the 1992~Sep~13 observation was longer and lasted for more than 4\,h
({\tt ExpTime} $\approx$ 14.6\,ks).

We complemented the temporal information above with that provided by the {\em
ROSAT} Mission web page at the Max-Planck-Institut f\"ur extraterrestrische
Physik (MPE)\footnote{\tt
http://www.mpe.mpg.de/xray/wave/rosat/index.php?lang=en}. 
The time spanning between start and end of daily observations were in general
larger than {\tt ExpTime}. 
For example, total pointing time in 1995~Feb~26 was 2.707\,ks, but the HRI
detector was switched off in three occasions for a summed time interval
of 0.342\,ks, thus giving {\tt ExpTime} = 2.365\,ks.
However, only in a few cases total pointing time was significantly (by a factor
at least two) larger than {\tt ExpTime}. 

All pointing coordinates were identical.
Since the {\em post facto} attitude determination accuracy of {\em ROSAT} was
6\,arcsec, the overlapping field of view between different pointings was maximal
(i.e. the full 38\,arcmin-size square centered on $\sigma$~Ori~AF--B).
Therefore, all sources were observed (or at least pointed at) in each of the
observations.
Besides, the processing site of all images was the MPE, which ensures the
homogeneity of the dataset.

\subsection{Optical/near-infrared counterparts}

We identified 24 groups of 1RXH events, each of them having more than 20
associated 1RXH events, $N$ (i.e. $N >$ 20). 
Due to the clear radial concentration of events (see below) and small size
($\rho <$ 10\,arcsec) of each group in comparison to their mutual separation
($\rho >$ 40\,arcsec), we found no problem to associate a group of 1RXH events
to every single X-ray source.
For an accurate follow-up analysis, we only selected X-ray sources with more
than 20 events.
Two discarded {\em ROSAT} X-ray sources with lower number of events are
linked to the $\sigma$~Orionis stars \object{Mayrit~403090}
([W96]~4771--1038; $N$ = 15) and  \object{Mayrit~374056} ([W96]~4771--1075; $N$
= 15). 
Remaining discarded X-ray sources have 11 or less associated 1RXH events.

Then, we searched with Aladin for the optical and near-infrared counterparts of
the X-ray sources with $N >$ 20.
The basic properties of the 24 identified X-ray sources and their corresponding
optical/near-infrared counterparts are given in
Tables~\ref{table.xrayparameters} and~\ref{table.opticalcounterparts}.
All except one counterpart are $\sigma$~Orionis cluster members in the Mayrit
catalog (Caballero 2008b).
The cluster non-member is a galaxy with a very strong X-ray emission already
detected by the {\em Einstein Observatory}, 2E~1456 (Section~\ref{2e1456}).

\begin{figure}
\epsscale{1.12}
\plotone{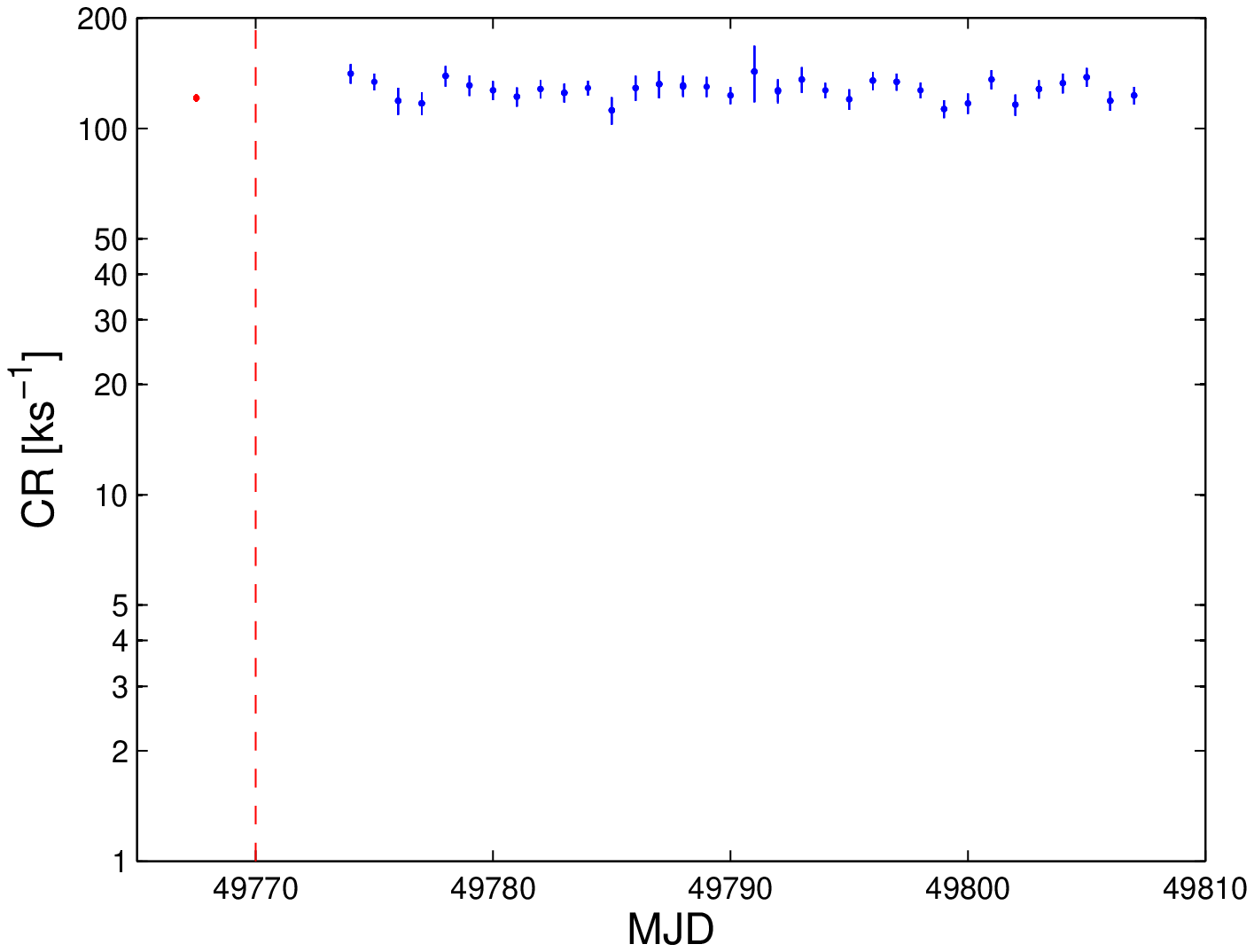}
\plotone{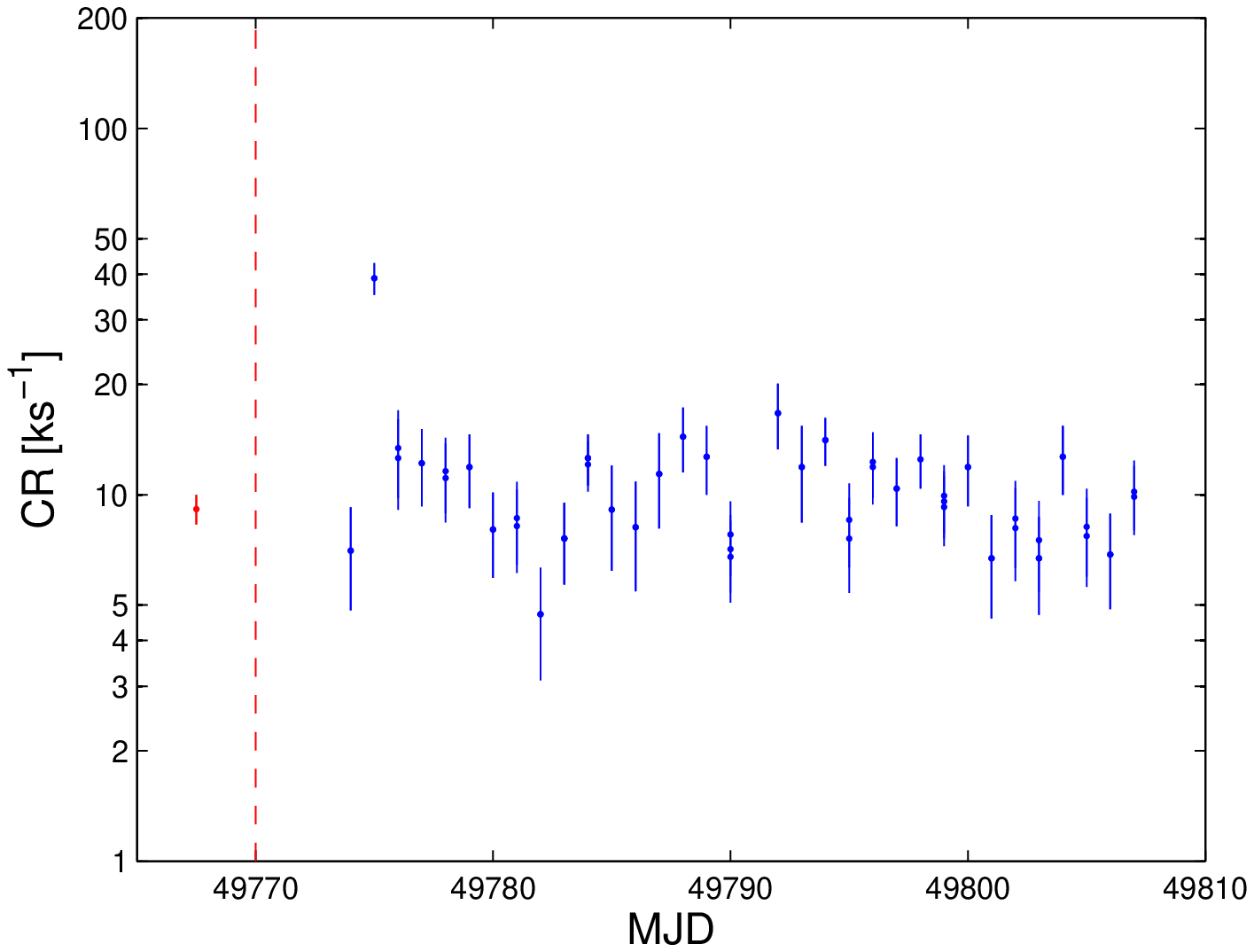}
\plotone{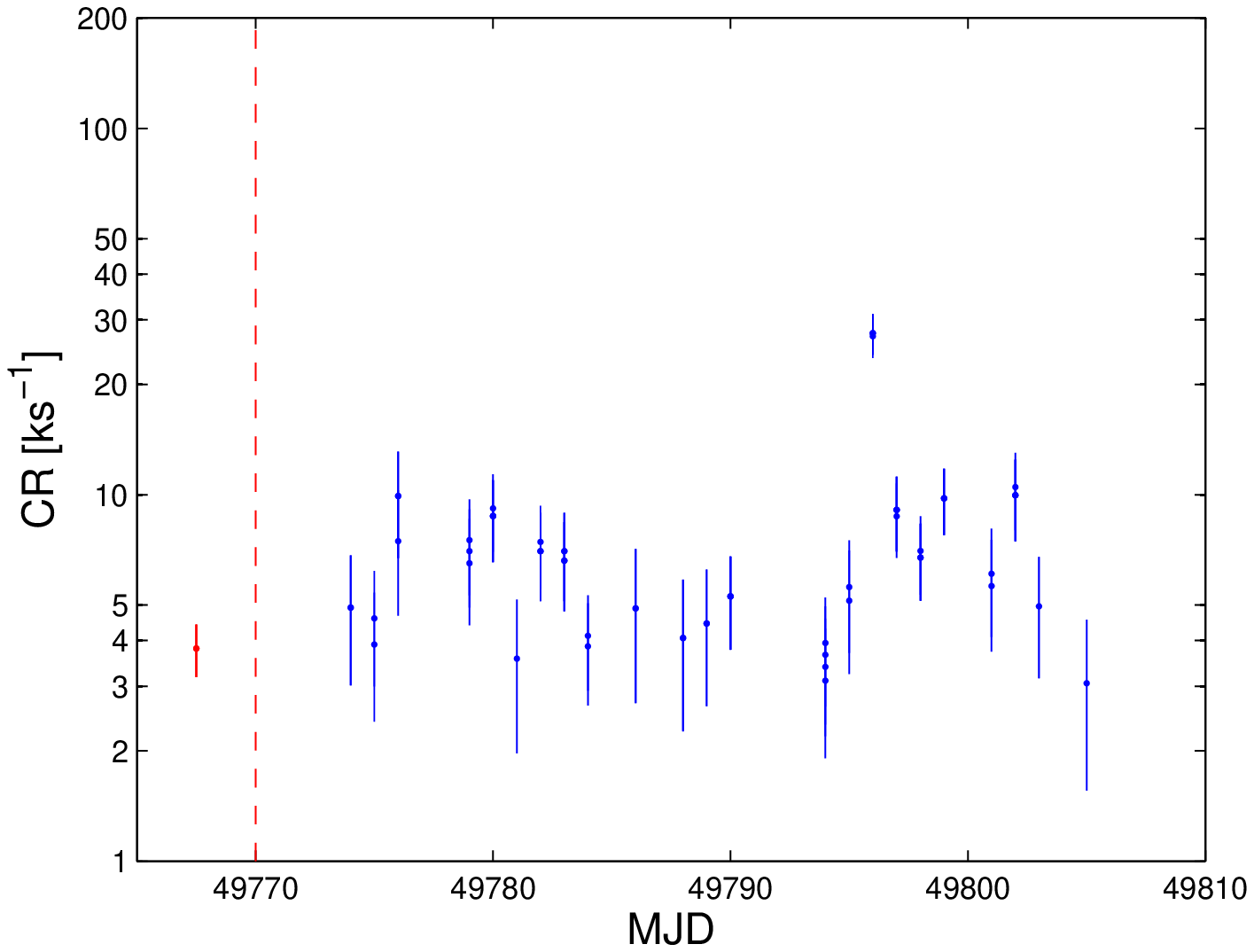}
\caption{Time series of the 1RXH events associated to $\sigma$~Ori~AF--B,
Mayrit~42062 ($\sigma$~Ori~E), and Mayrit~105249, from top to bottom.
The isolated data points to the left of the vertical dashed line correspond to
the 1992~Sep measurements;
we have conveniently shifted these data points in the horizonal axis for a
better visualization.
\label{fig.X1}}
\end{figure}

\begin{figure}
\epsscale{1.12}
\plotone{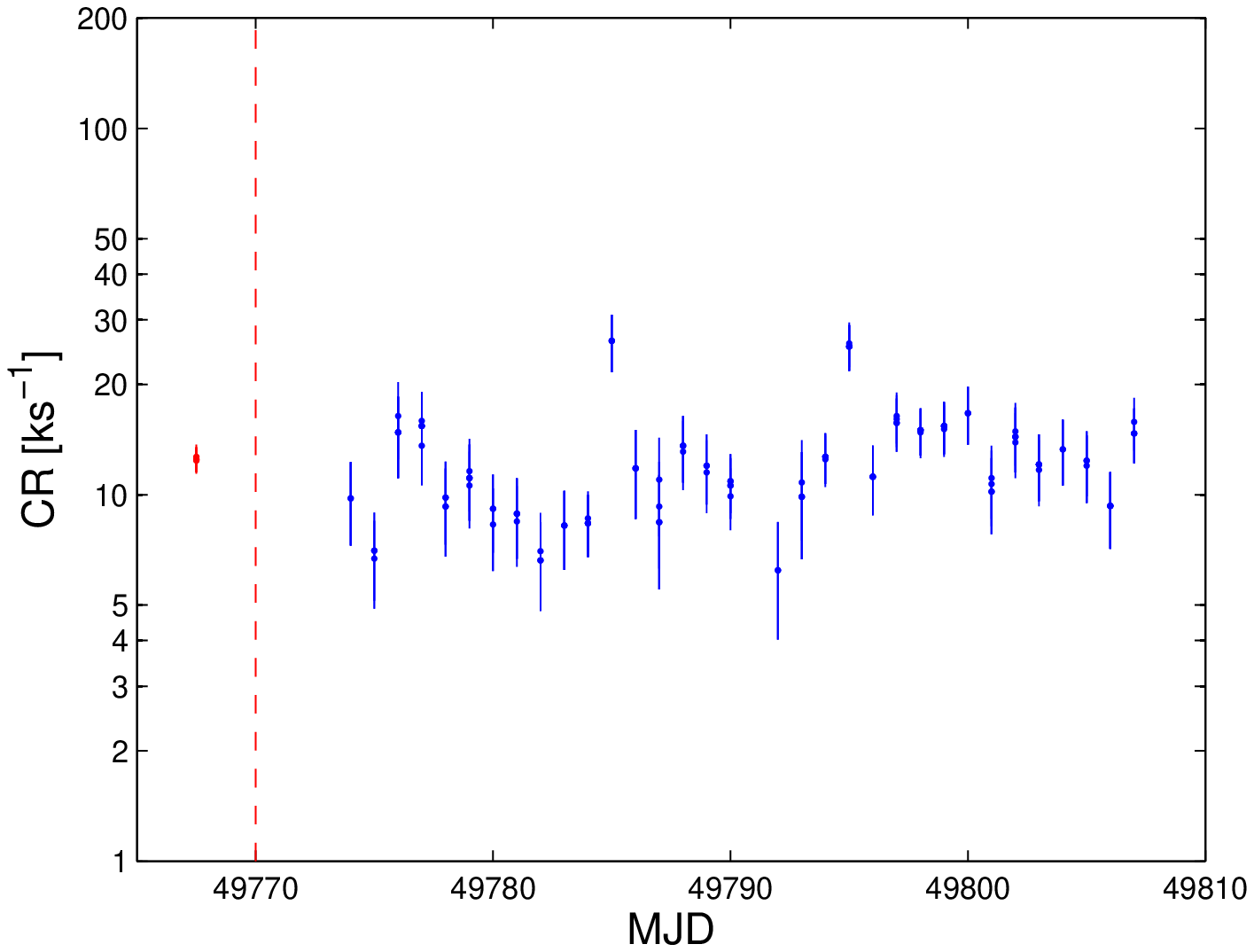}
\plotone{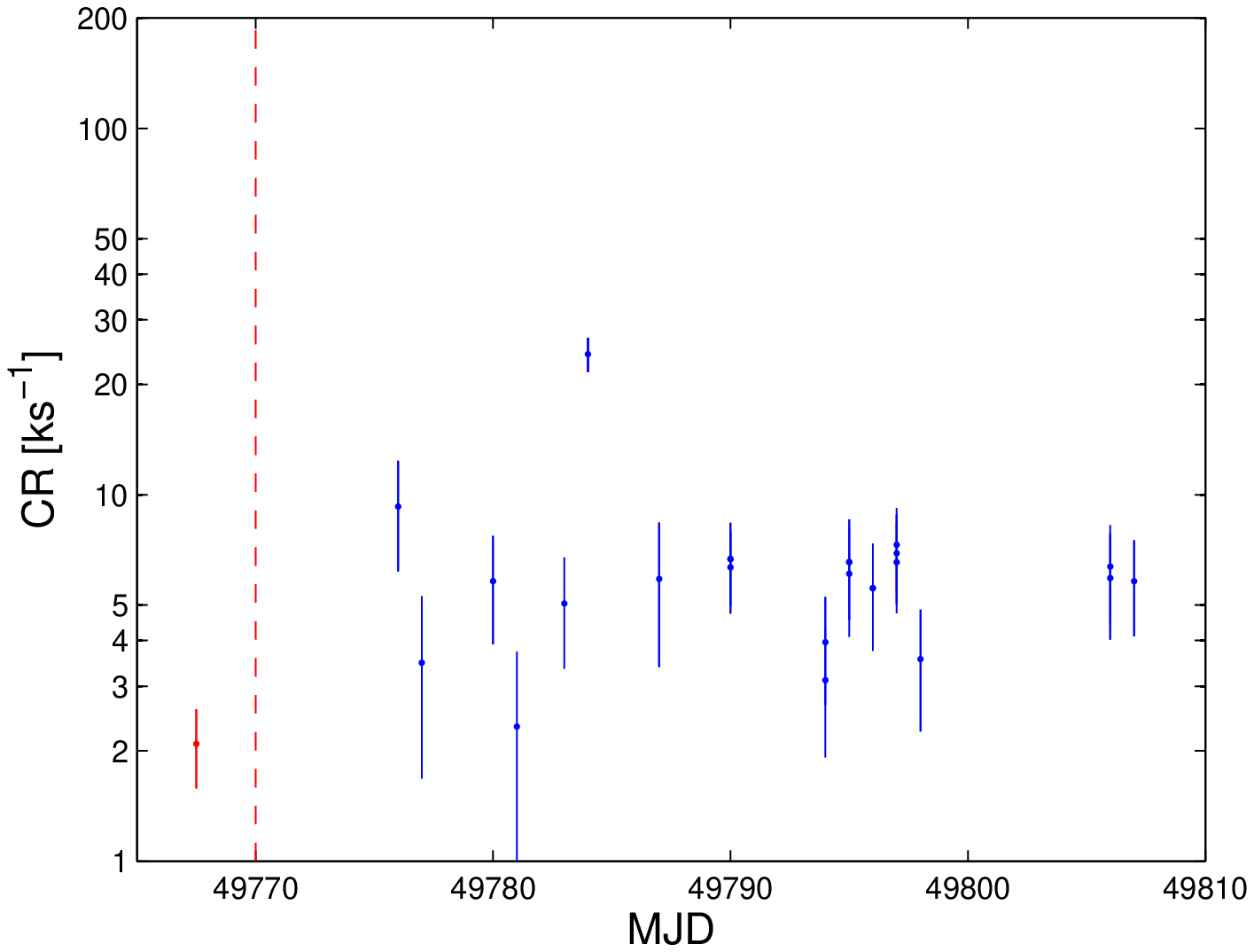}
\plotone{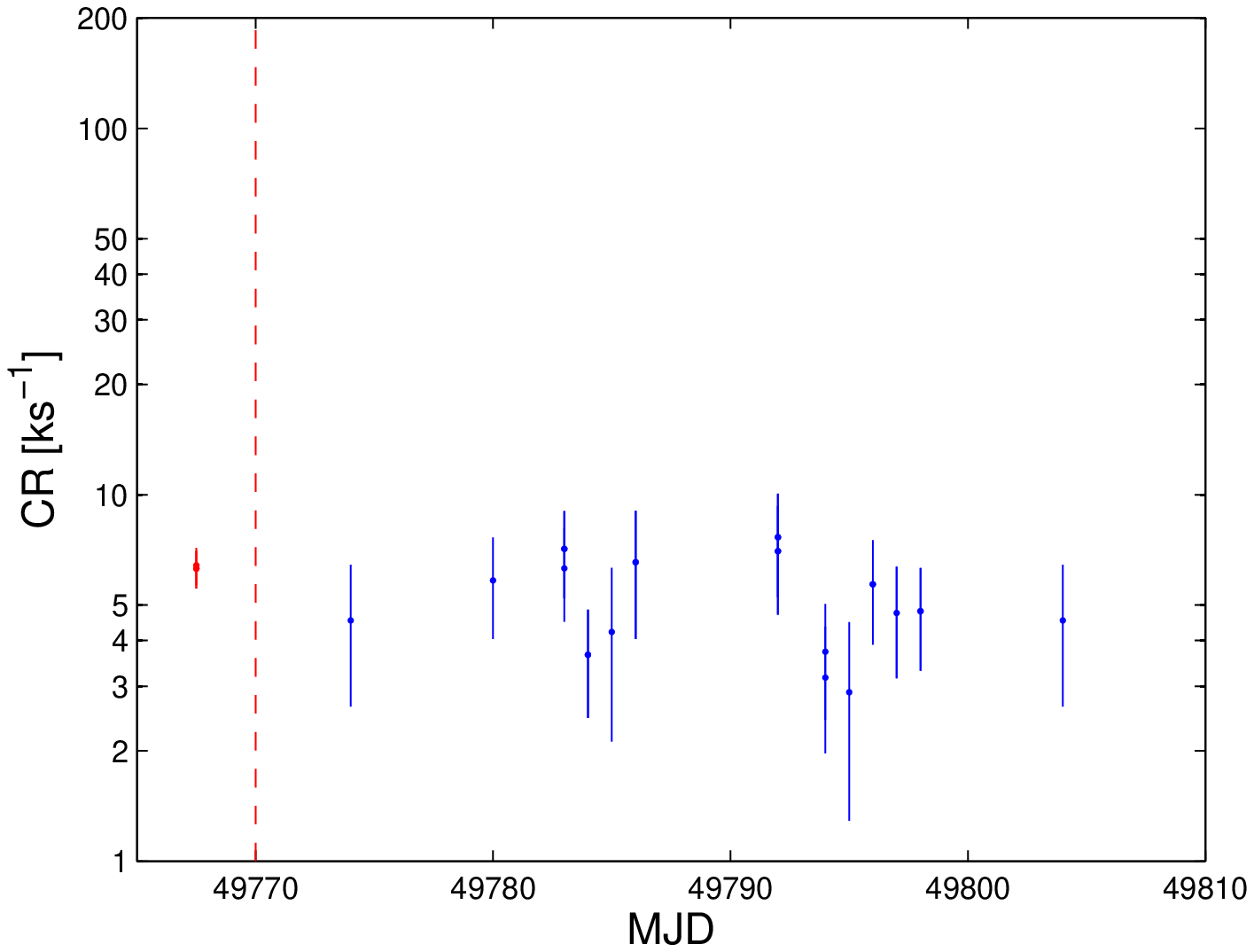}
\caption{Same as Fig.~\ref{fig.X1}, but for Mayrit~114305, Mayrit~156353, and
Mayrit~157155.
\label{fig.X2}}
\end{figure}

\begin{figure}
\epsscale{1.12}
\plotone{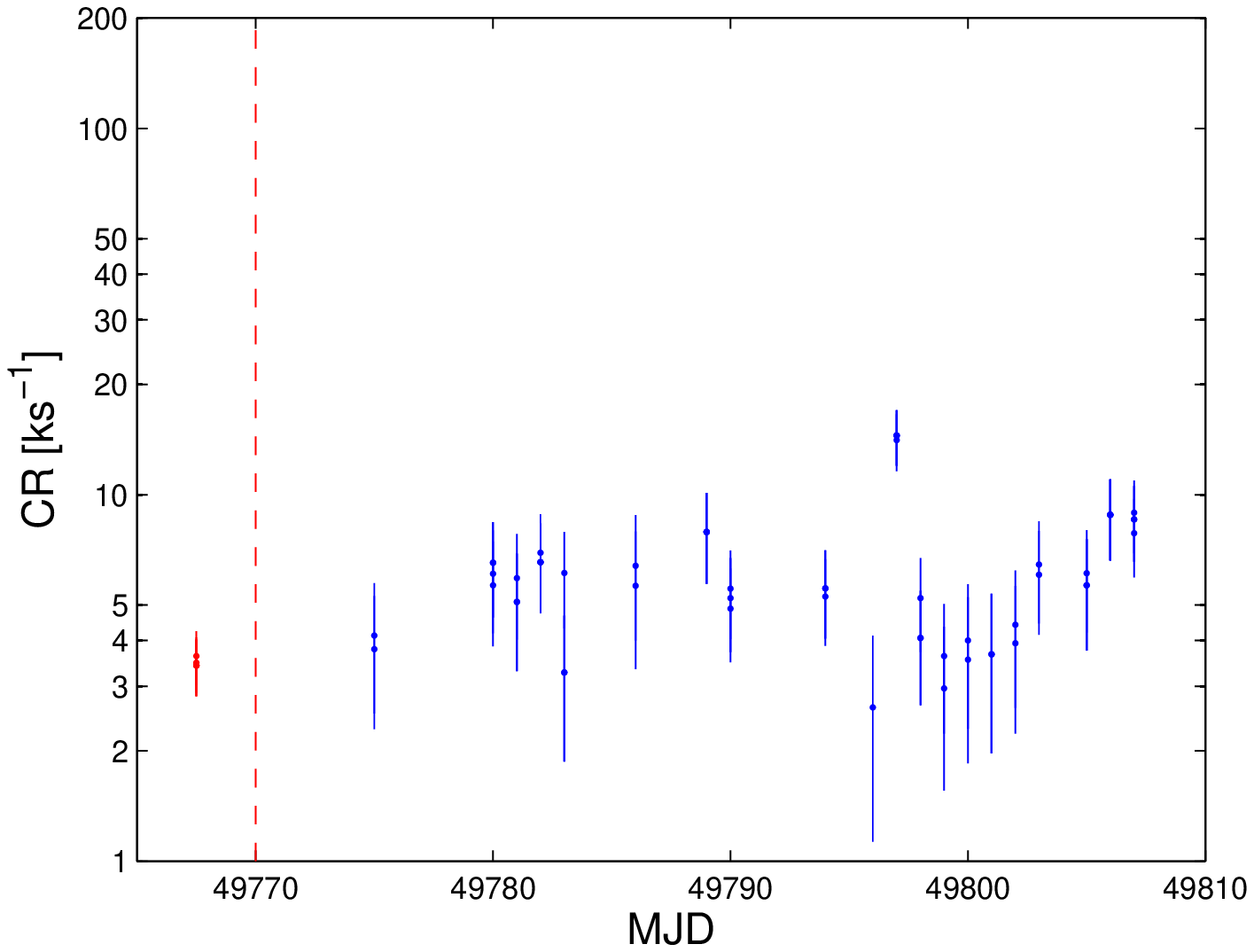}
\plotone{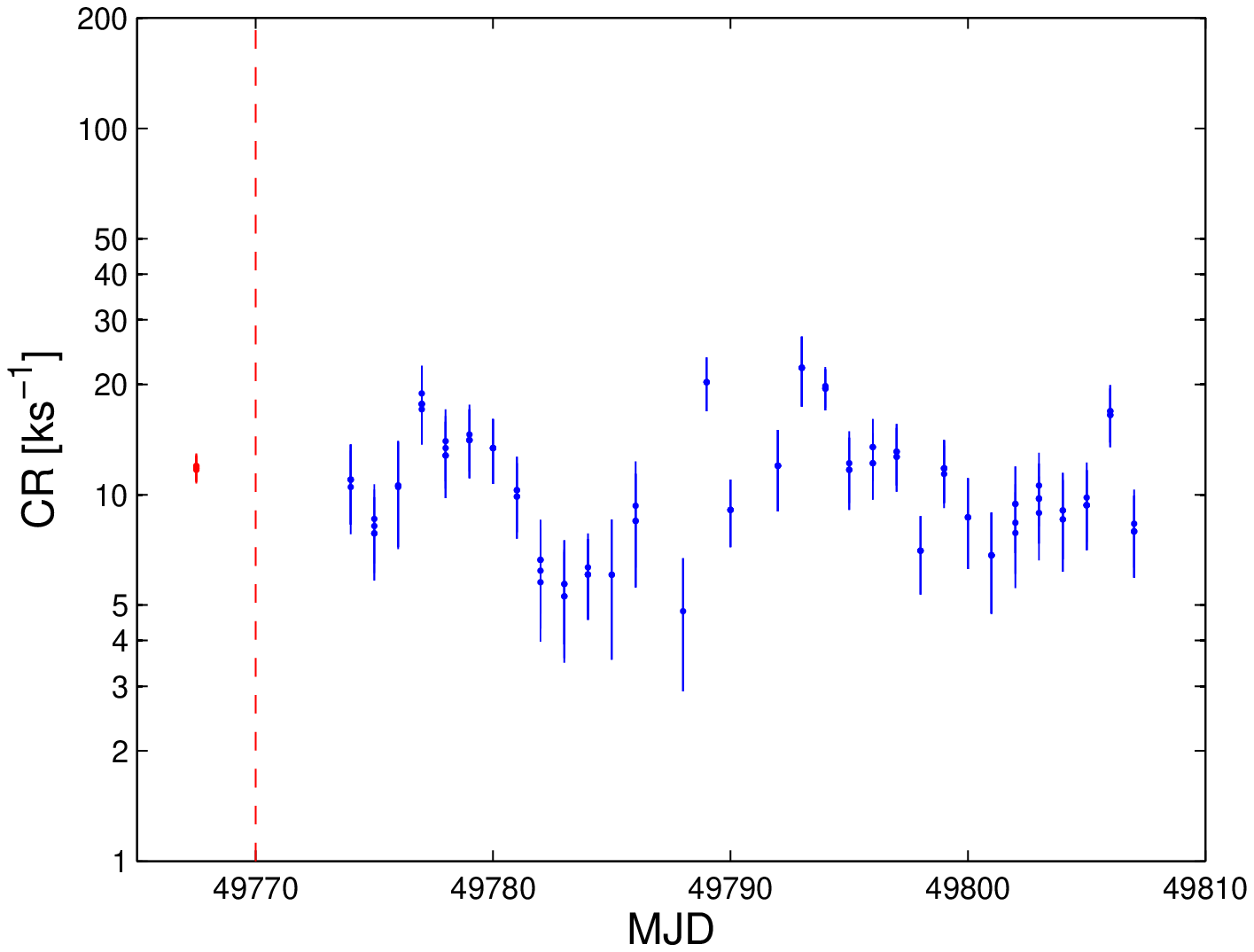}
\plotone{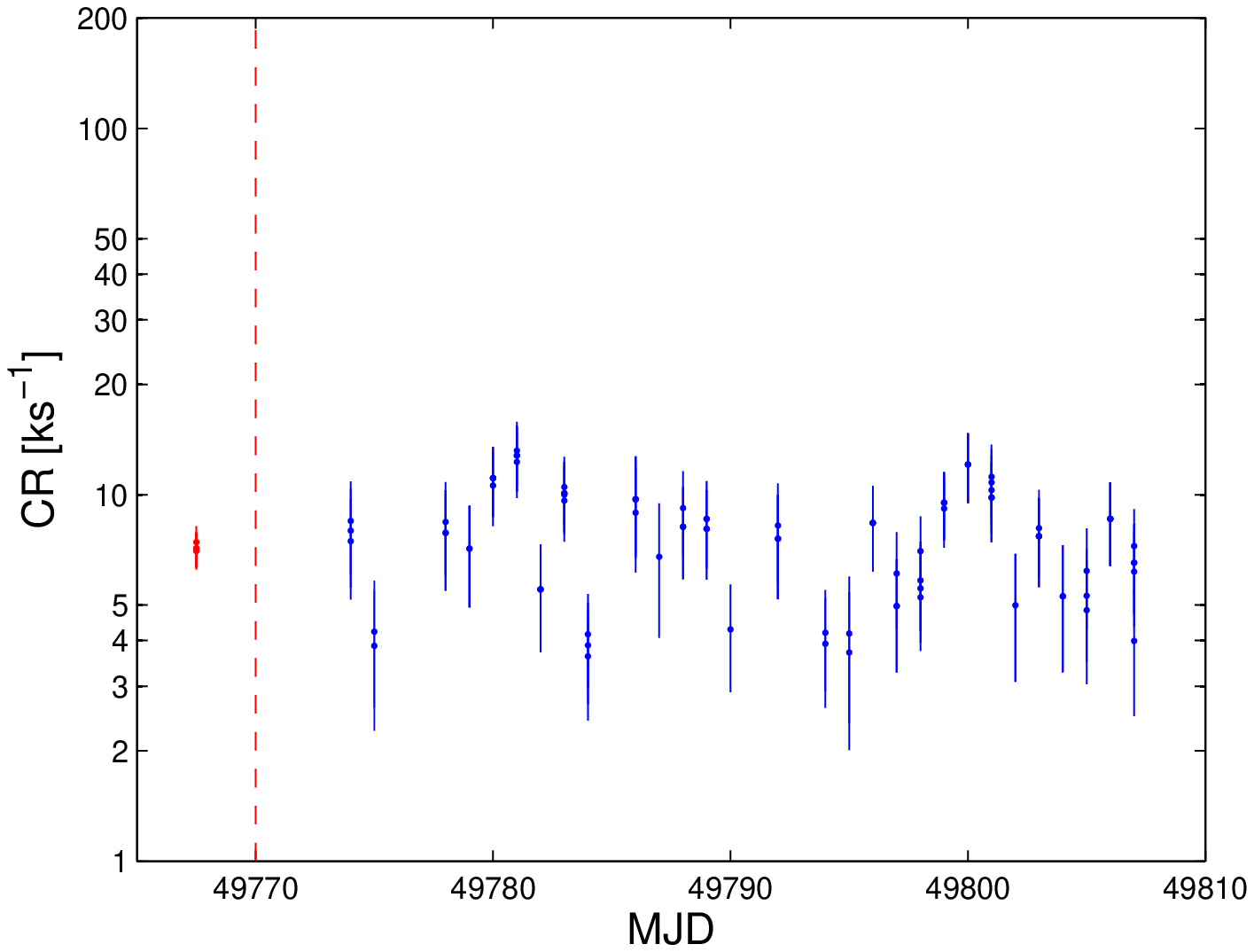}
\caption{Same as Fig.~\ref{fig.X1}, but for Mayrit~180277, Mayrit~203039, and
Mayrit~207358.
\label{fig.X3}}
\end{figure}

\begin{figure}
\epsscale{1.12}
\plotone{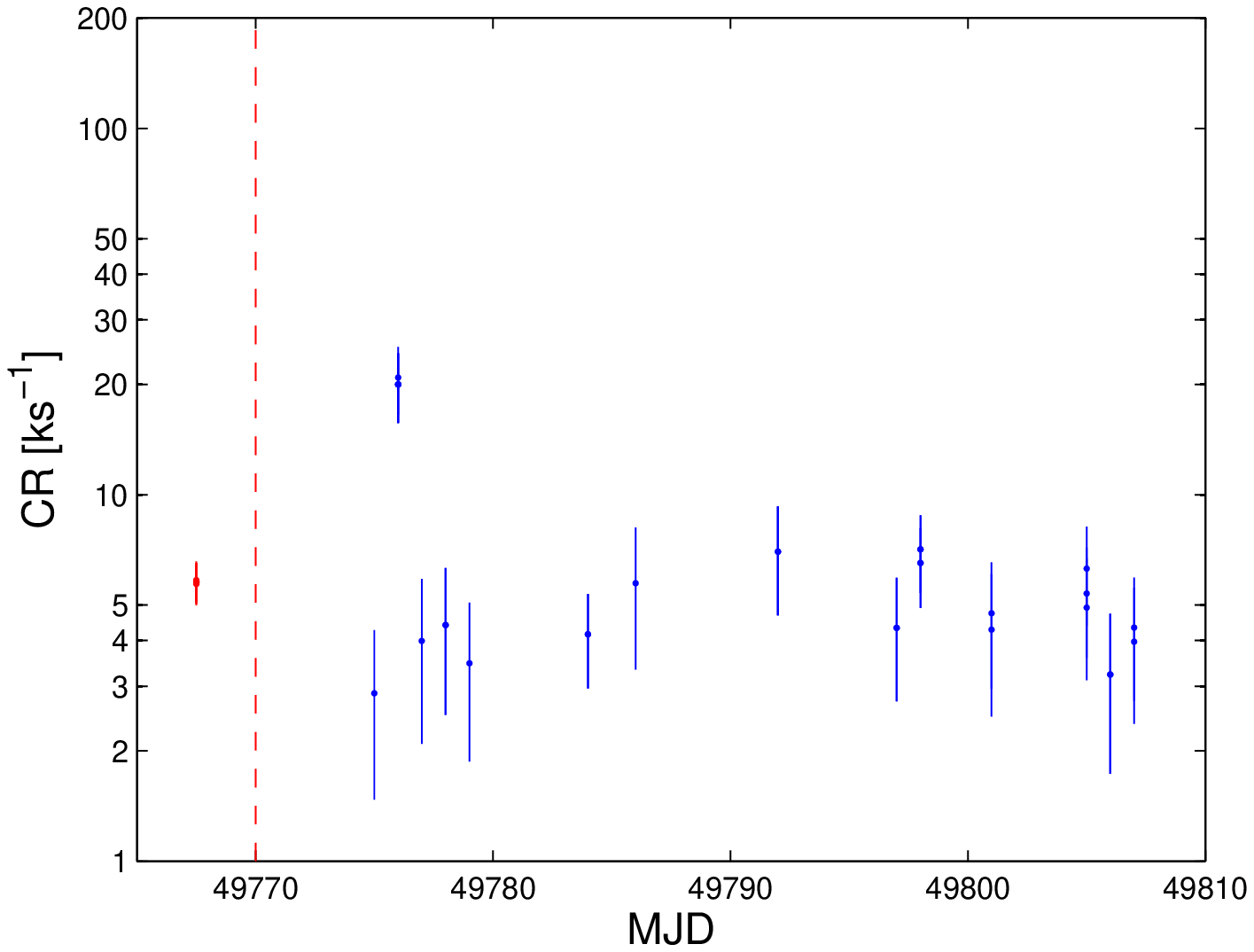}
\plotone{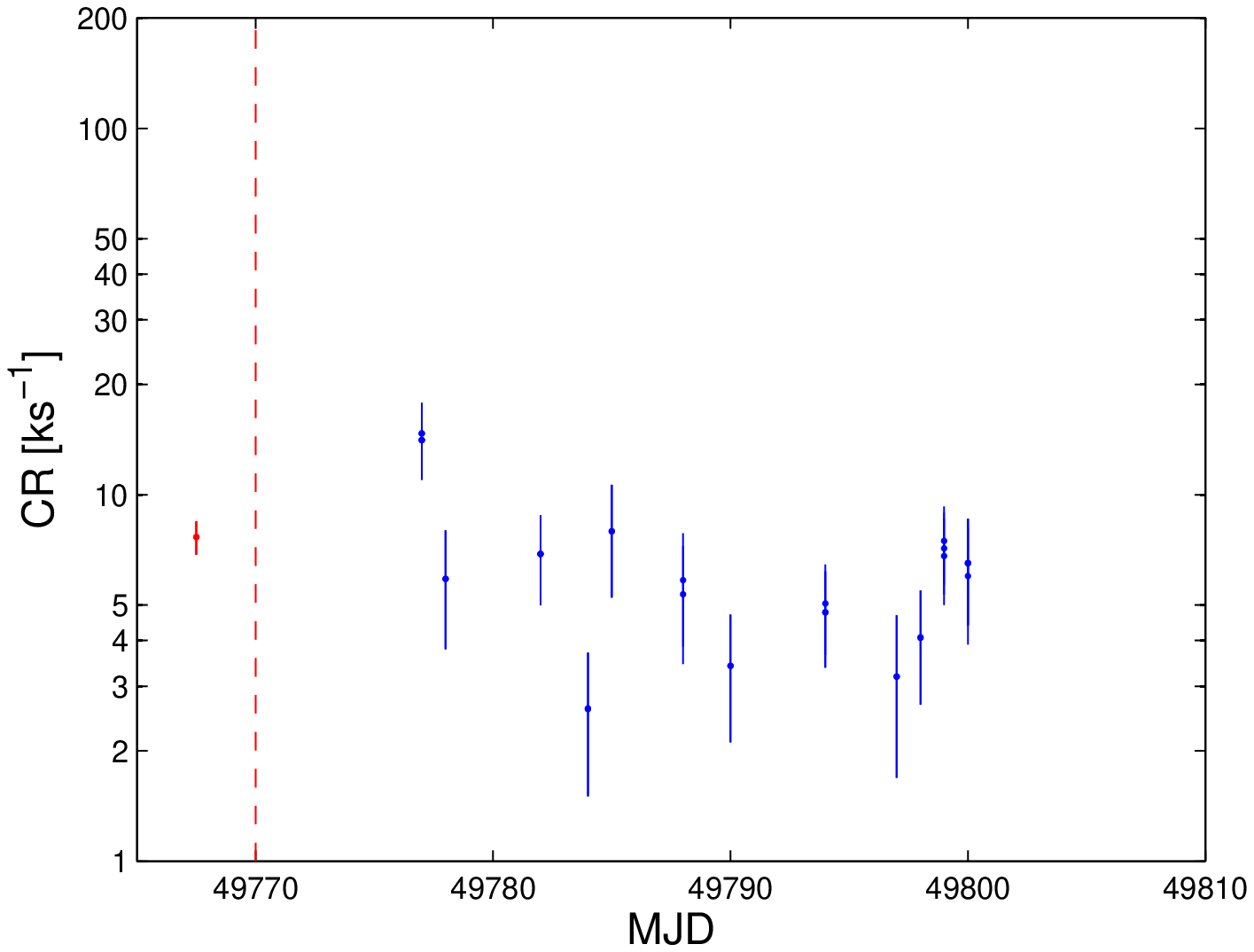}
\plotone{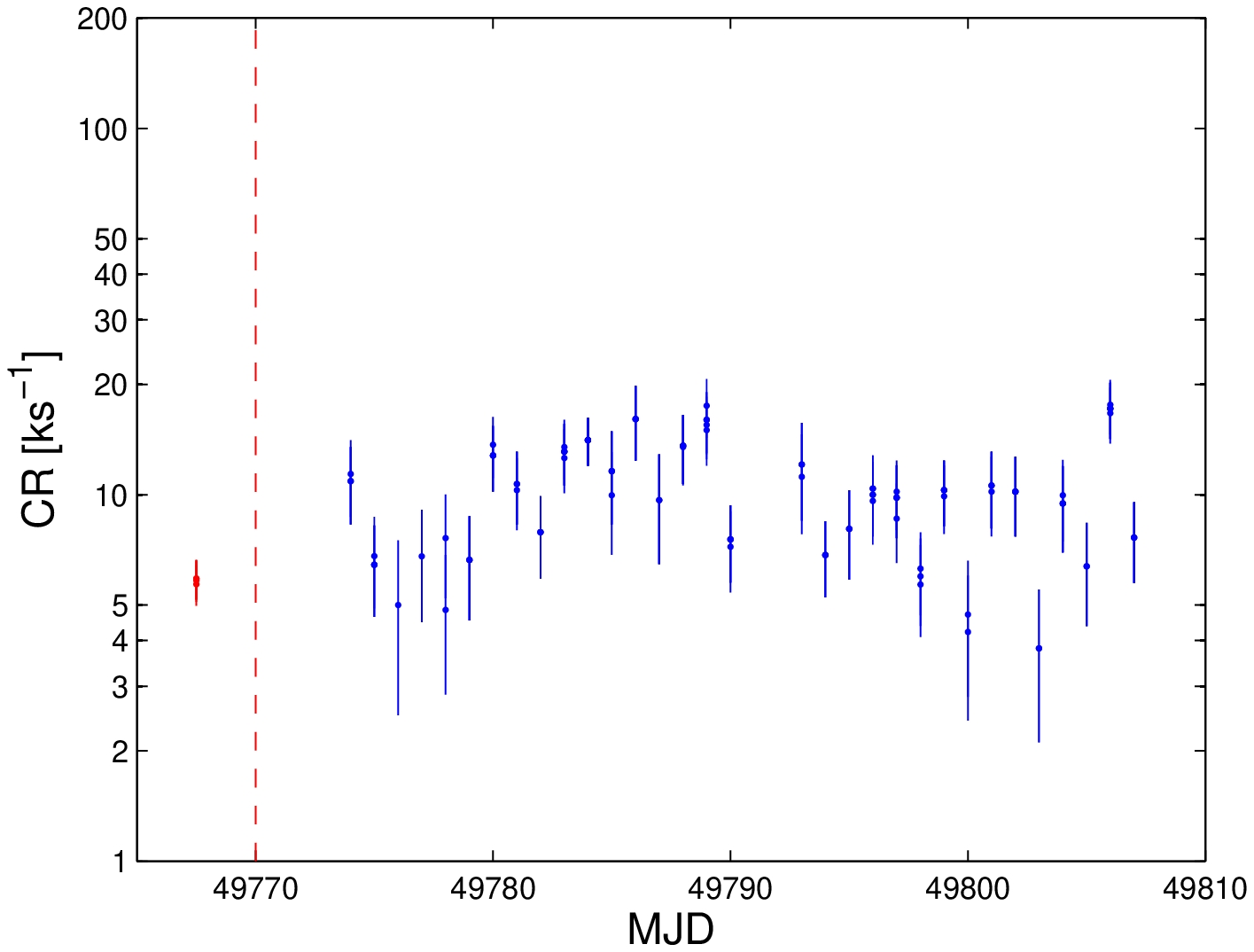}
\caption{Same as Fig.~\ref{fig.X1}, but for Mayrit~260182, Mayrit~285331, and
Mayrit~306125.
\label{fig.X4}}
\end{figure}

\begin{figure}
\epsscale{1.12}
\plotone{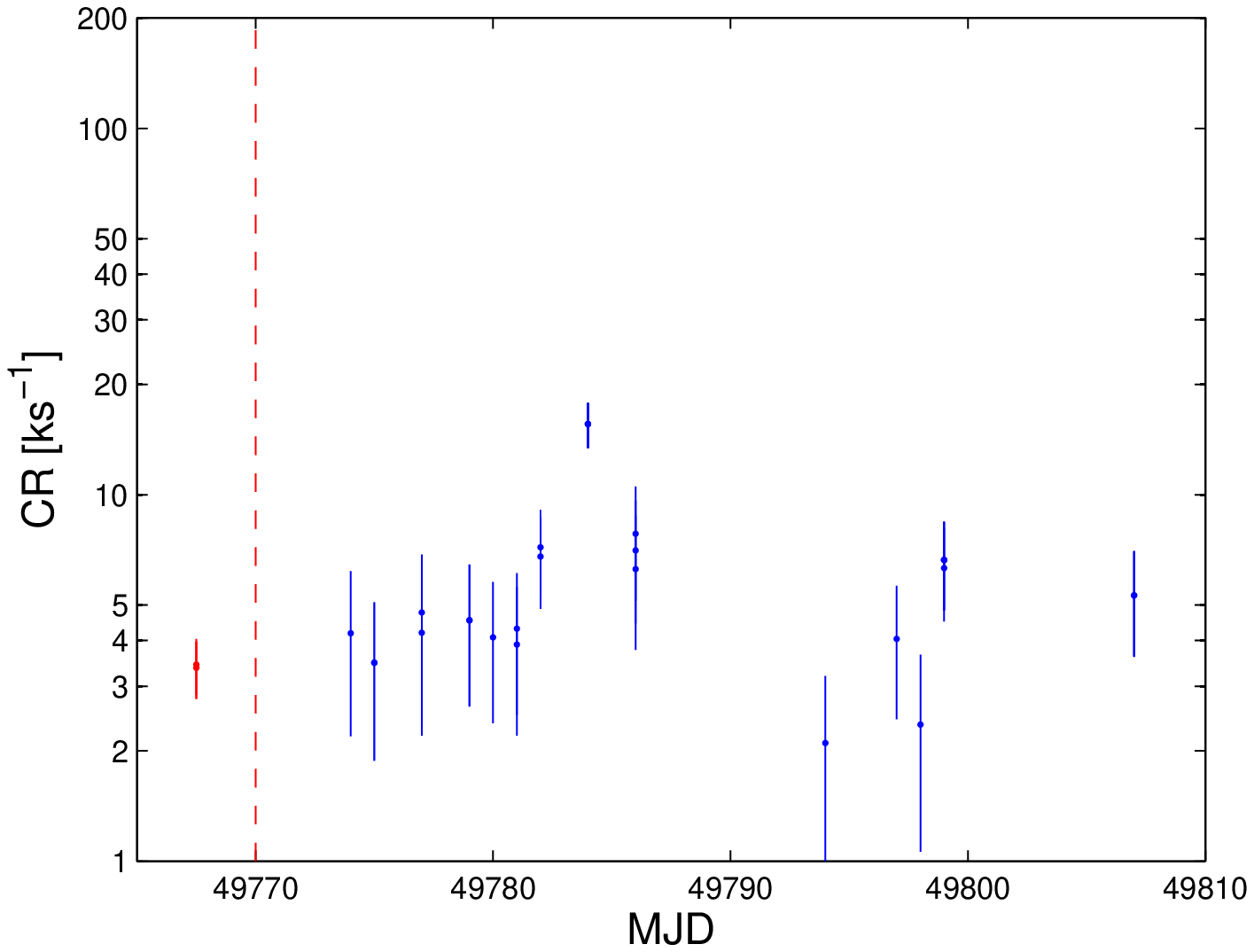}
\plotone{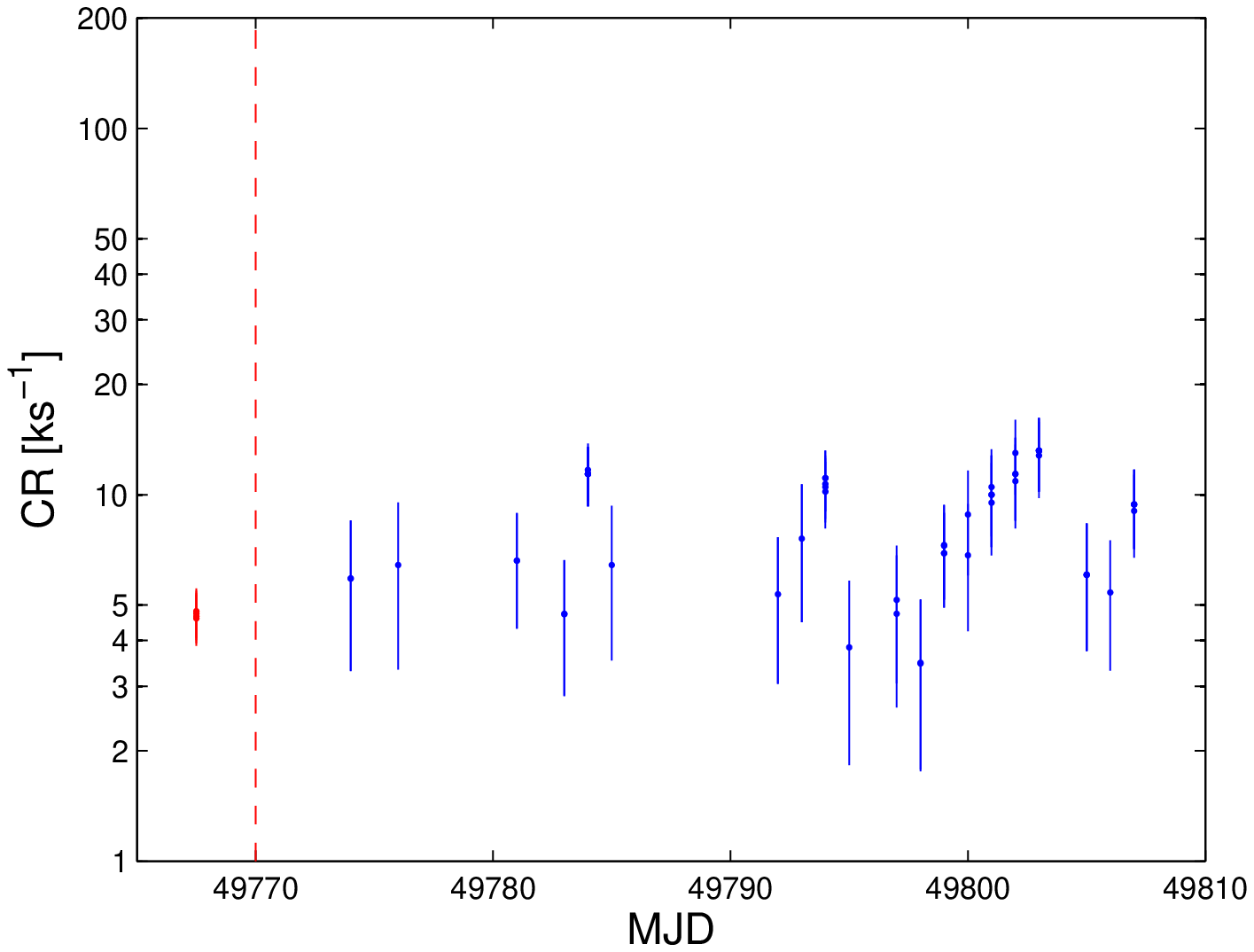}
\plotone{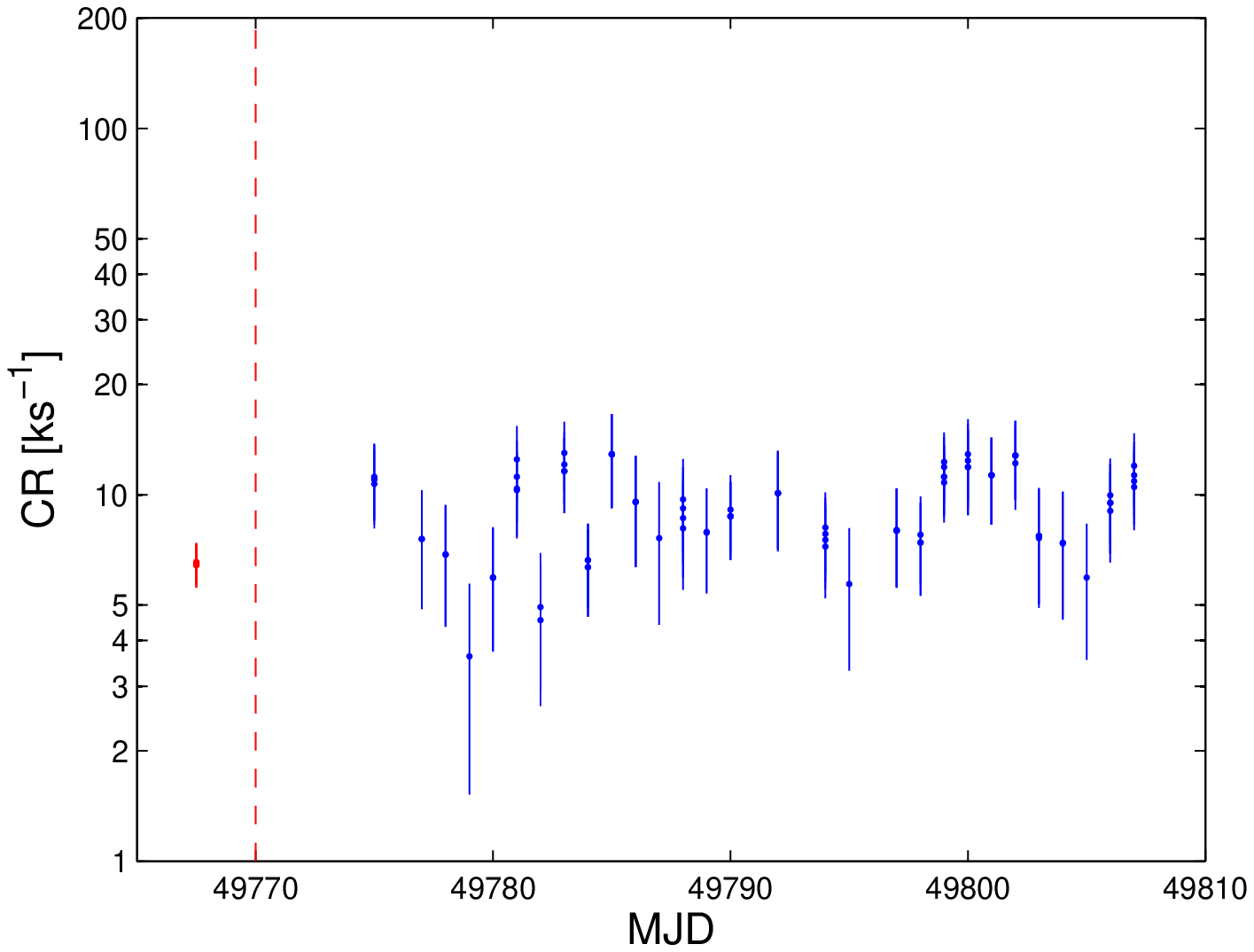}
\caption{Same as Fig.~\ref{fig.X1}, but for Mayrit~344337, Mayrit~528005, and
Mayrit~615296.
\label{fig.X5}}
\end{figure}

\begin{figure}
\epsscale{1.12}
\plotone{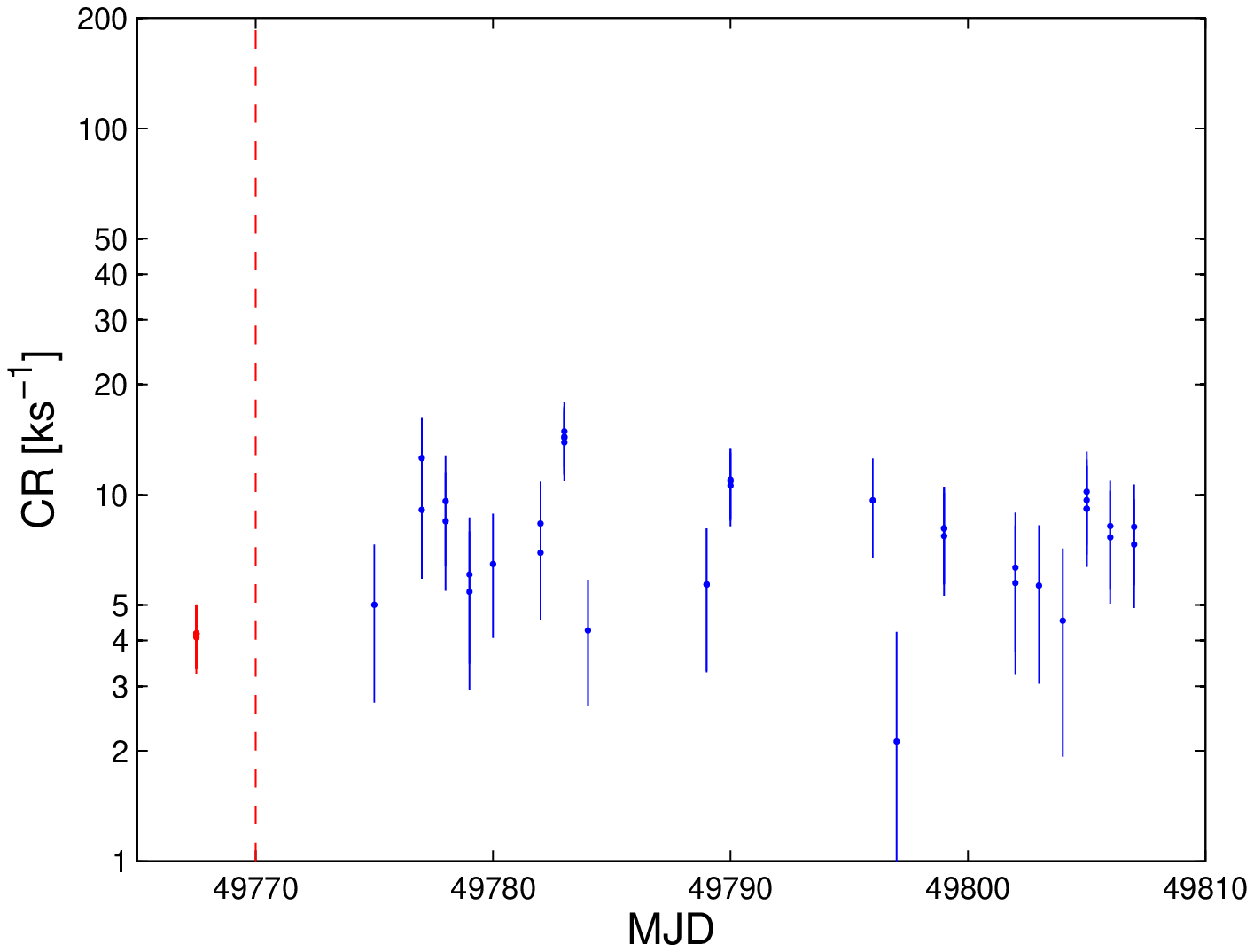}
\plotone{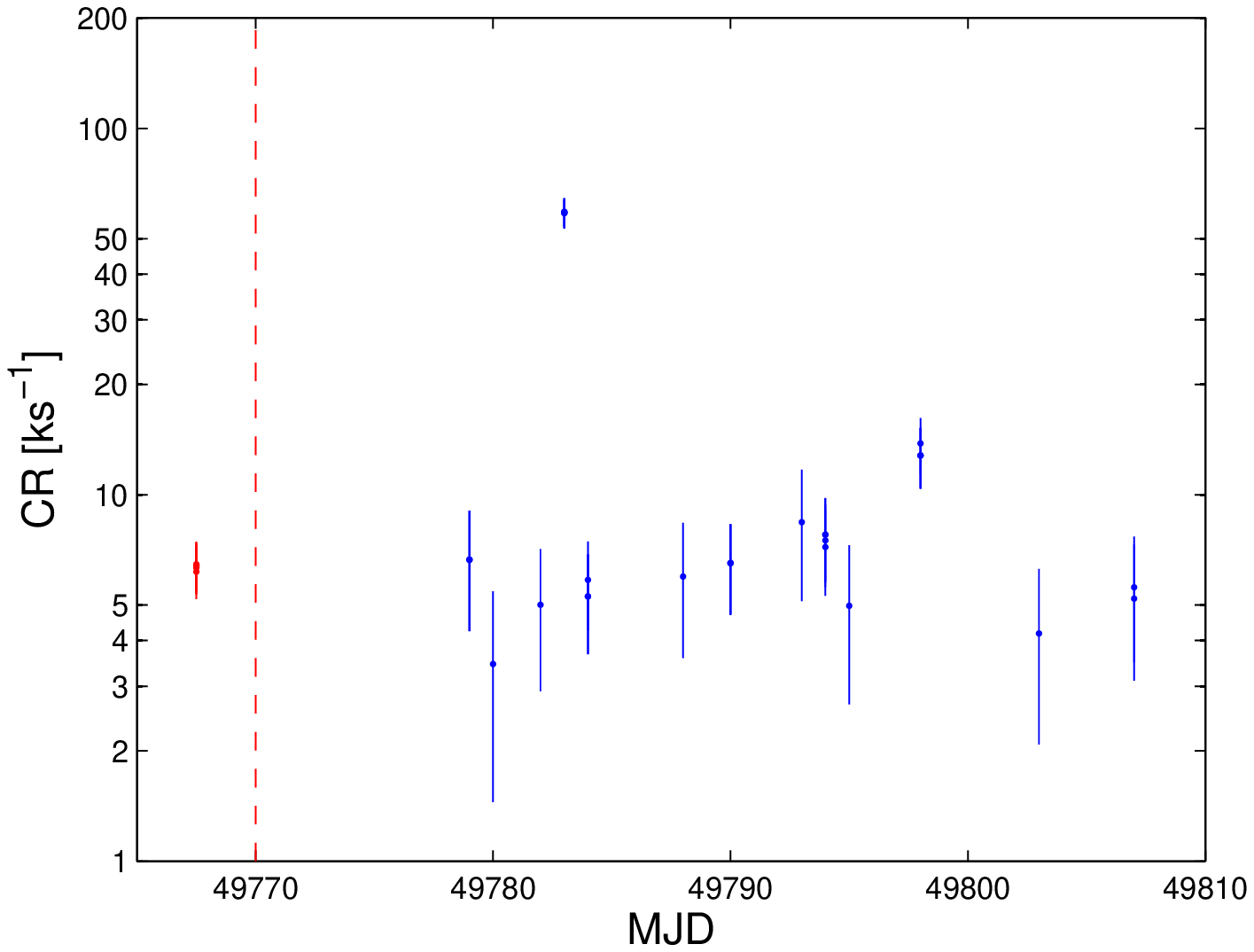}
\plotone{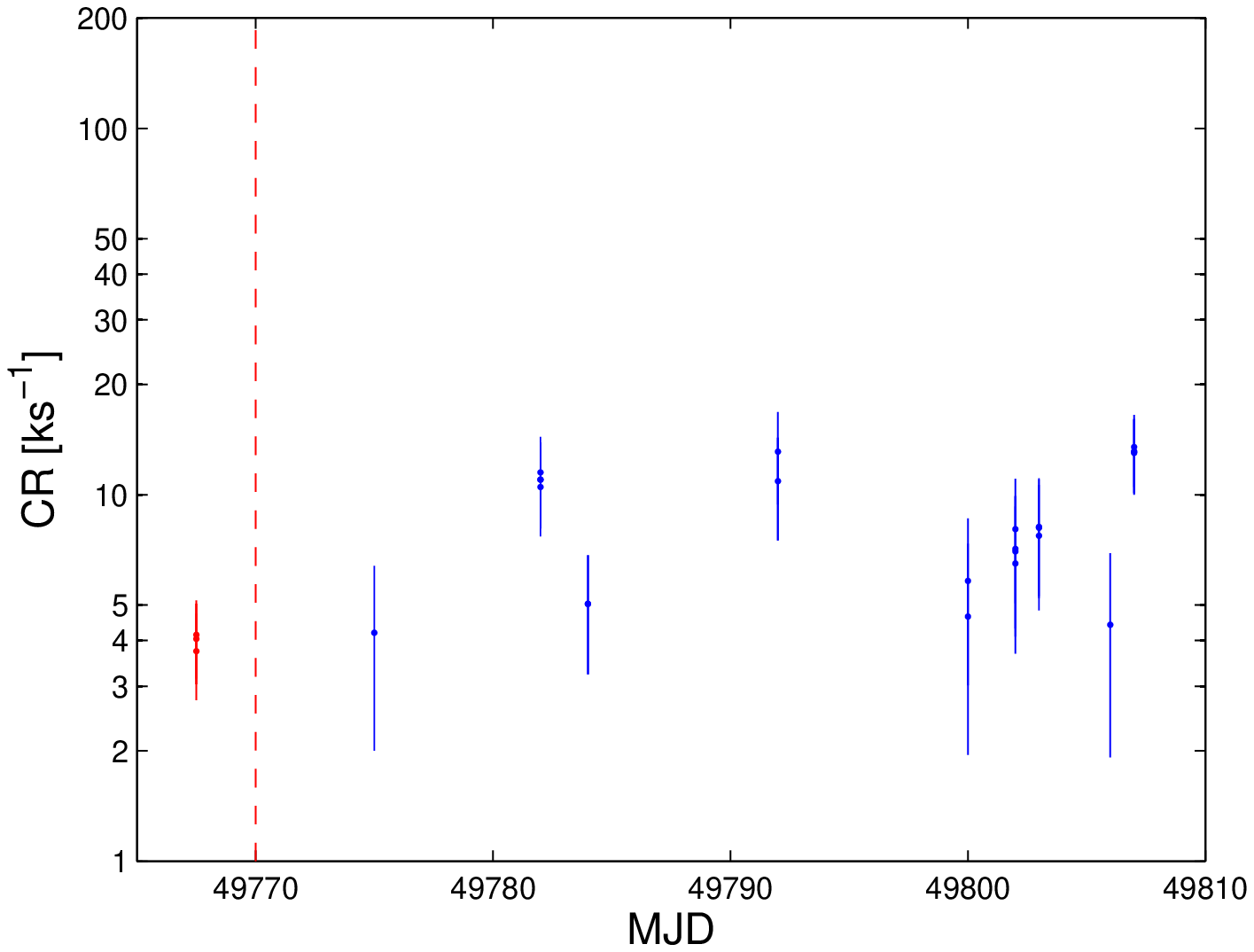}
\caption{Same as Fig.~\ref{fig.X1}, but for Mayrit~634052, Mayrit~653170, and
Mayrit~750107.
\label{fig.X6}}
\end{figure}

\begin{figure}
\epsscale{1.12}
\plotone{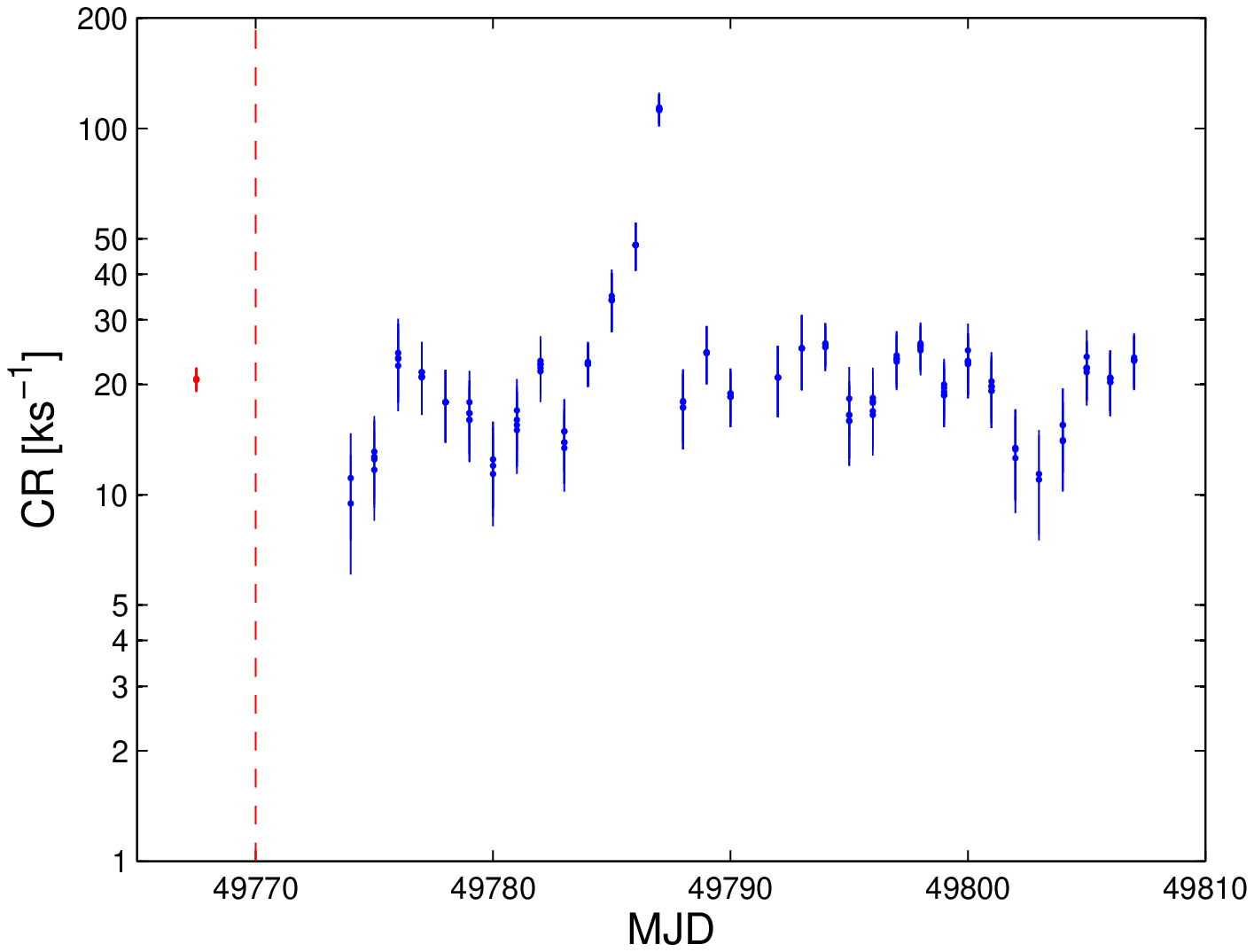}
\plotone{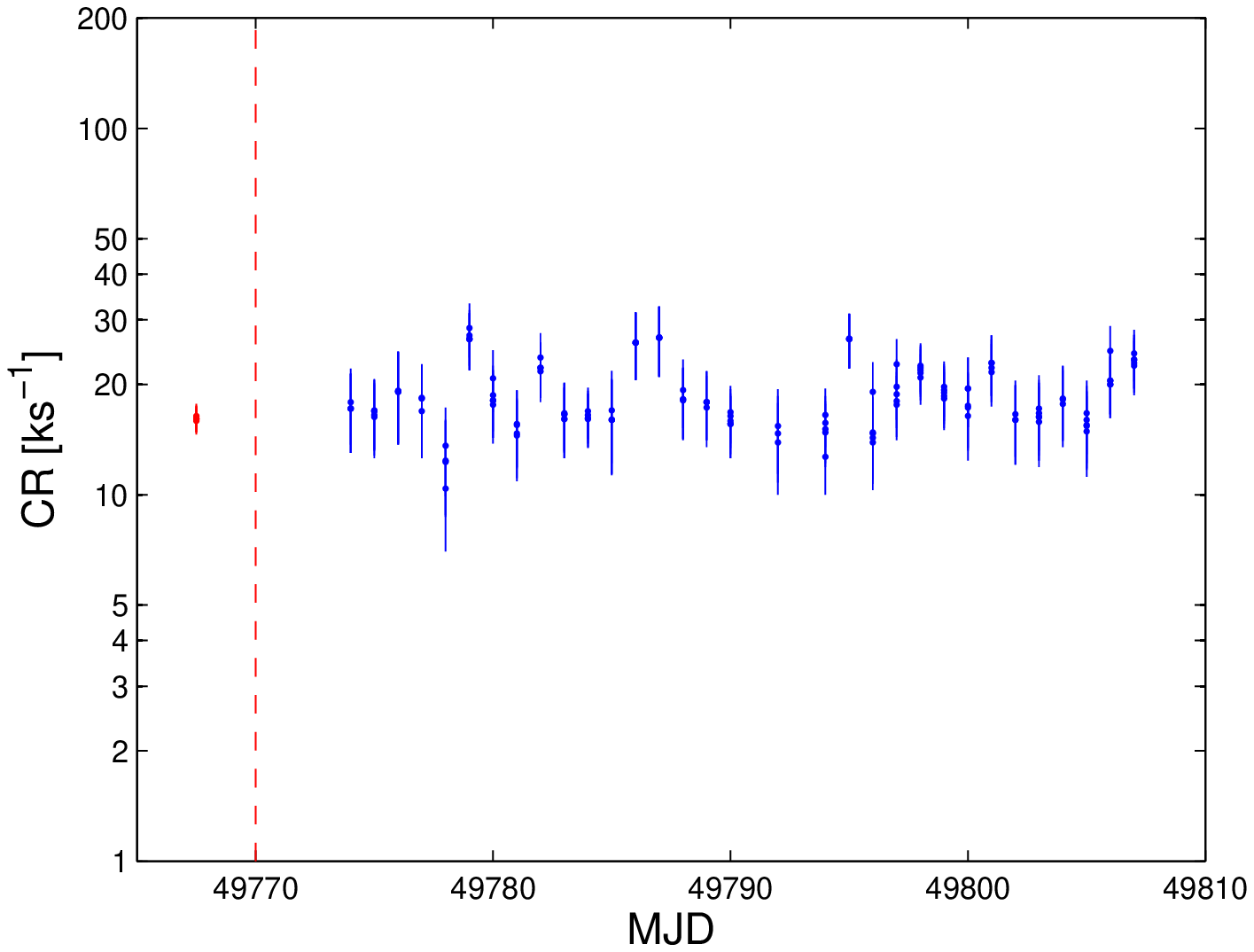}
\plotone{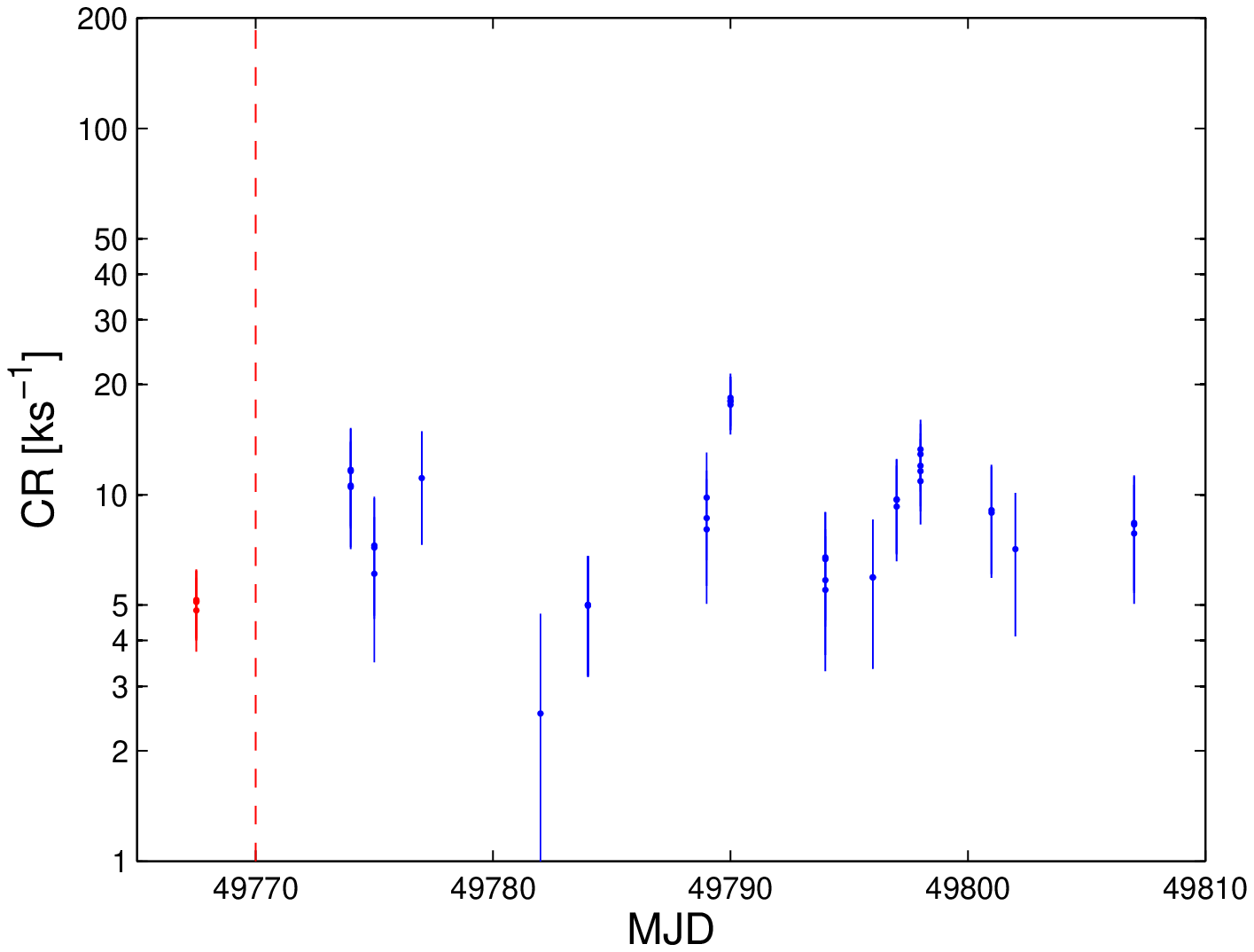}
\caption{Same as Fig.~\ref{fig.X1}, but for Mayrit~783254, Mayrit~789281, and
Mayrit~822170.
\label{fig.X7}}
\end{figure}

\begin{figure}
\epsscale{1.12}
\plotone{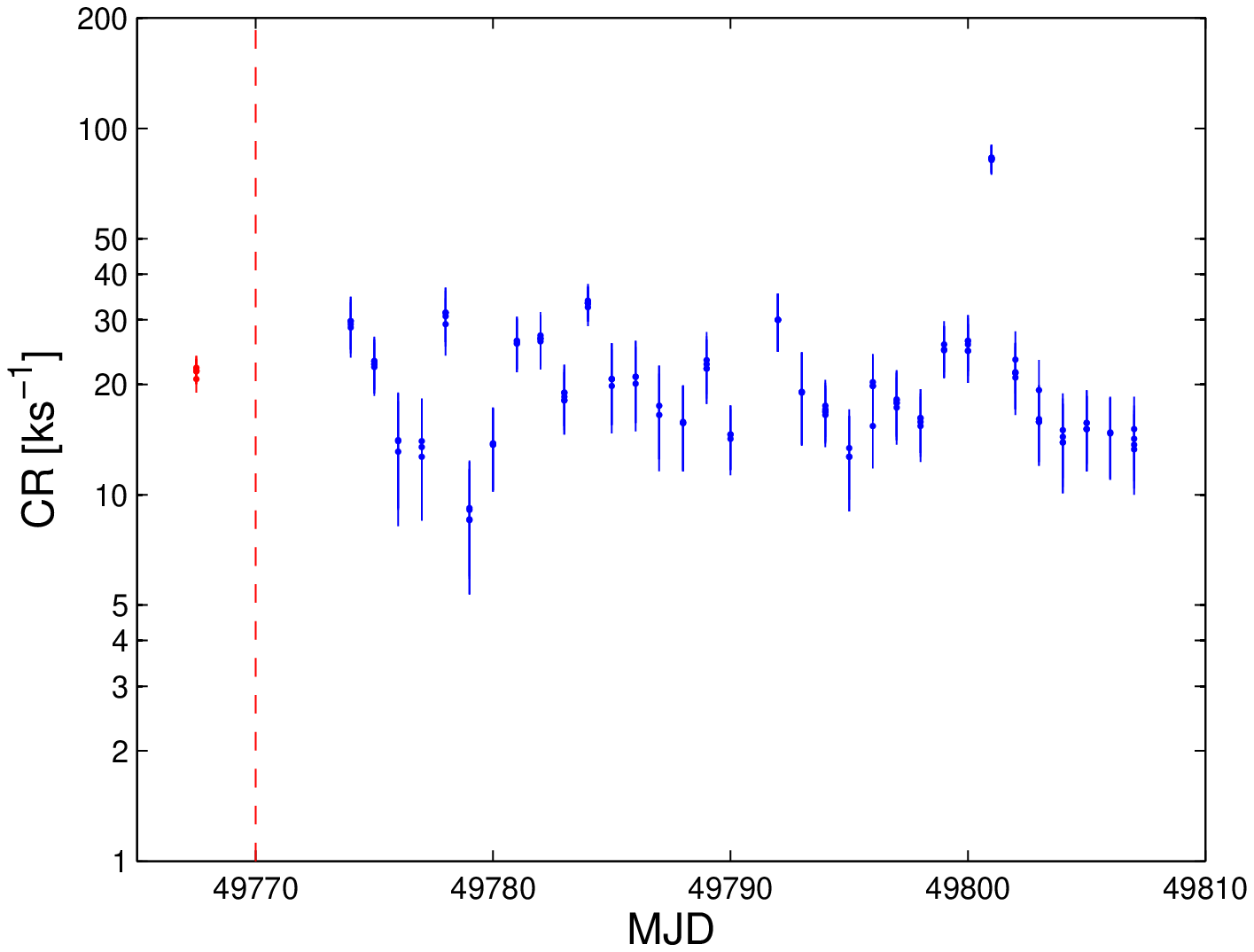}
\plotone{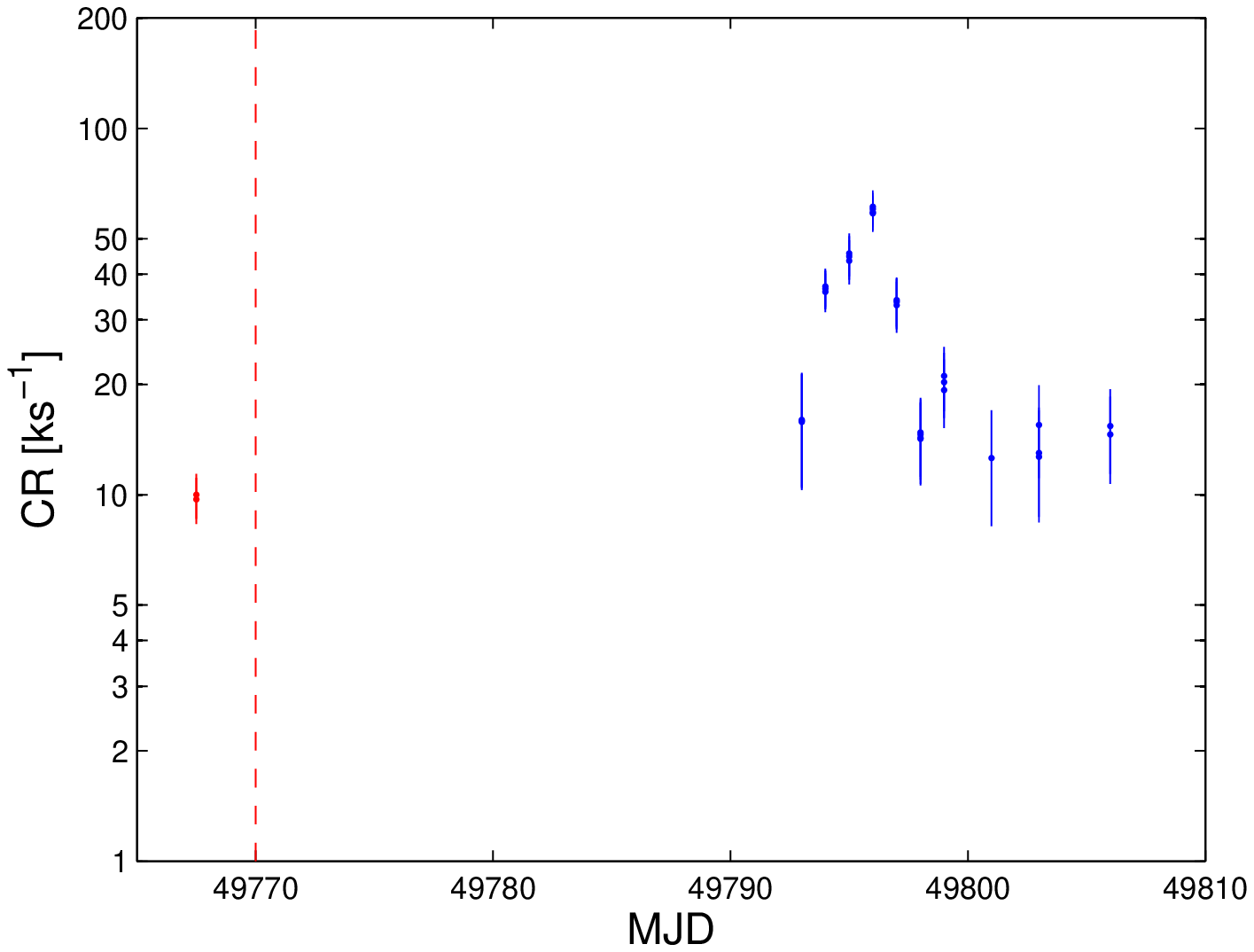}
\plotone{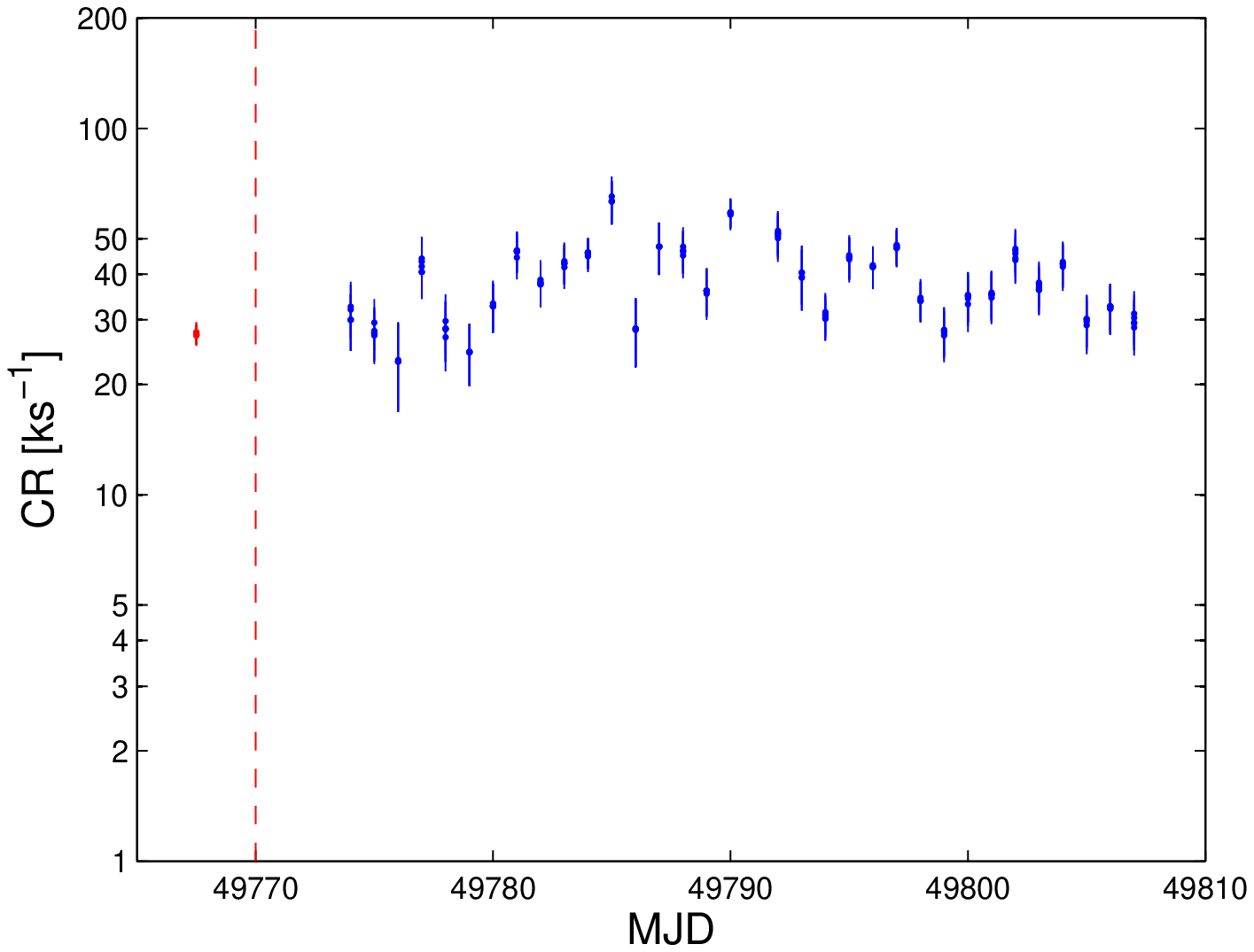}
\caption{Same as Fig.~\ref{fig.X1}, but for Mayrit~863116, Mayrit~969077, and
galaxy 2E~1456.
\label{fig.X8}}
\end{figure}

On the one hand, in Table~\ref{table.xrayparameters} we provide the means and
standard deviations of the right ascension and declination ($\overline \alpha$,
$\overline \delta$, $\sigma_\alpha$, $\sigma_\delta$), number of associated
events ($N$), mean and standard deviation of the net count rate
($\overline{CR}$, $\sigma_{CR}$), and mean of the error on the count rate
($\overline{\delta CR}$) of each group of 1RXH events associated to an X-ray
source. 
In the antepenultimate column, we list a normalized double-weighted
$\chi^2$ of the X-ray series computed as in Fuhrmeister \& Schmitt (2003; see
Section~\ref{xrayvariability})\footnote{Note the non-standard definition of
$\chi^2$ and the different nomenclature in Fuhrmeister \& Schmitt (2003) for
the same parameters: $\overline{CR} \equiv \mu$, $CR_i \equiv r_i$, $\delta CR_i
\equiv e_i$.  
Besides, in their work, the number of events and of scans was identical ($N
\equiv n$).}:  
\begin{equation}
\chi^2 = \frac{1}{N} \sum \frac{(\overline{CR}-CR_i)^2}{\delta CR_i^2}.
\label{eq.chi2} 
\end{equation}
\noindent The time series (or ``light curves'') of the 24 X-ray sources are
shown in Figs.~\ref{fig.X1} to~\ref{fig.X8}.
For the typical length of an individual observation, {\tt Exptime}
$\approx$ 2.2\,ks, the faintest sources with $\overline{CR}$ = 5--10\,ks$^{-1}$
have corresponding low total (source + background) counts within the radius of
source circle for intensity measurement, of about 10--20 counts in each of the
short observations.
In occasions, these sources were too faint even to be detected.
Likewise, there may be several events associated to a bright X-ray source for
each daily observation ($\dot{N} \sim$ 4.8\,d$^{-1}$ for the brightest ones). 

On the other hand, in Table~\ref{table.opticalcounterparts} we provide the
recommended alternative names, Two-Micron All Sky Survey (2MASS) coordinates and
$J$-band magnitudes, and remarks from the literature for the
optical/near-infrared counterparts of the X-ray sources in
Table~\ref{table.xrayparameters}.
Deep Near Infrared Survey of the Southern Sky (DENIS) $i$ and 2MASS $HK_{\rm
s}$ magnitudes of the sources were tabulated in the Mayrit catalog
(Caballero~2008b). 

As can be seen in Table~\ref{table.xrayparameters}, the accuracy of the
astrometry of the X-ray sources, quantified by the standard deviations
$\sigma_\alpha$ and $\sigma_\delta$, varies with the distance to the instrument
axis (i.e. the center of the field of view) from $\sim$1.5\,arcsec for
$\sigma$~Ori~AF--B to $\sim$4.8\,arcsec for Mayrit~969077 (at $\rho \approx$
969\,arcsec $\approx$ 16.2\,arcmin, $\theta \approx$ 77\,deg, to
$\sigma$~Ori~AF--B). 
This variation is due to the well-known degradation of the point spread
function with the distance to the instrument axis of {\em ROSAT} in particular
and of all X-ray missions in general.
In any case, our spatial resolution of 2--4\,arcsec after averaging is
comparable to that one of a single {\em XMM-Newton} pointing in
$\sigma$~Orionis (Watson et~al. 2003, 2009; Franciosini et~al. 2006;
L\'opez-Santiago \& Caballero 2008).

\begin{figure}
\epsscale{1.12}
\plotone{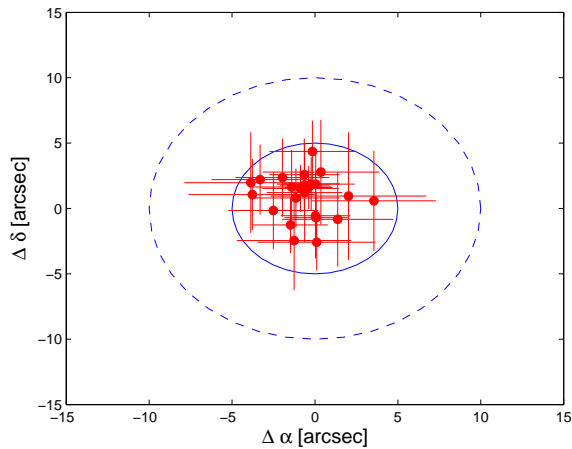}
\caption{Separations between the 24 X-ray sources and their respective
optical/near-infrared counterparts ($\Delta \delta$ vs. $\Delta \alpha$
diagram).
Solid and dashed lines mark deviations of 5 and 10\,arcsec, respectively.
Actual maximum separations are about 4\,arcsec. 
\label{fig.DraDde}}
\end{figure}

We are confident of our identification of optical counterparts.
The separation between the X-ray sources and their respective optical
counterparts, illustrated in Fig.~\ref{fig.DraDde}, is null within uncertainties
($\Delta \alpha = -0.7 \pm 1.7$\,arcsec, $\Delta \delta = +0.9 \pm
1.7$\,arcsec).
All of them except Mayrit~750107 had been tabulated by Caballero (2008b) as
strong X-ray emitters (with marks ``XX'' or~``XXX'').

While several of the X-ray sources had been already detected by the {\em
Einstein Observatory} (with 2E designation), most of them, including
Mayrit~750107, had been followed up with {\em ROSAT} by Wolk (1996; with [W96]
designations), {\em XMM-Newton} by Franciosini et~al. (2006; with [FPS2006]
designation), and {\em Chandra} by Skinner et~al. (2008). 
However, some strong X-ray sources and optical counterparts at the largest
angular separations from the $\sigma$~Orionis center, where most of X-ray
studies have focused on, had escaped from previous {\em XMM-Newton} and {\em
Chandra} observations (e.g. Mayrit~969077 had been identified only by the
{\em Einstein Observatory}).

Of the 23 Mayrit stars, three are (multiple) OB-type systems and 15 have a
fundamental spectroscopic feature of youth: Li~{\sc i} $\lambda$6707.8\,{\AA}
in evident absorption.
Some of them have also H$\alpha$ $\lambda$6562.8\,{\AA} in strong, asymmetric
emission, which is an indication of accretion from a disk, or even have flux
excesses at 3.8--8.0\,$\mu$m detected with {\em Spitzer} (marked with
``class~II'' in Table~\ref{table.opticalcounterparts}; Hern\'andez et~al. 2007
also claimed that Mayrit~114305 and Mayrit~634052 have evolved and debris disks,
respectively). 
The remaining five Mayrit stars without spectroscopic features of youth are
strong X-ray sources with stellar spectral energy distributions in Franciosini
et~al.~(2006) and colors and magnitudes consistent with membership in cluster.

\subsection{X-ray variability}
\label{xrayvariability}

Among the huge number of variability indicators that exist in the literature,
we used a novel method based on the comparison of the measured $\chi^2$ 
(Eq.~\ref{eq.chi2}) of our X-ray series and a large ensemble of
Monte~Carlo $\chi^2$s of simulated, non-variable, X-ray series. 
In short, we classified X-ray sources as variable if their $\chi^2$s are
much larger than those of stable sources.
Since the number of stable sources in the field of view was expected to be low
(Feigelson \& Montmerle 1999 hypothesized that all X-ray emitters, especially 
young stellar objects, are X-ray variable to some degree), we relied on
simulating a large number of suitable light curves of ``comparison stars''.
As defined in Eq.~\ref{eq.chi2}, our $\chi^2$ depends on the number of events,
$N$, associated to each X-ray source, the mean net count rate, $\overline{CR}$,
and the individual count rates and errors, $CR_i$ and $\delta CR_i$.
For a correct comparison, the simulated X-ray series must share identical
parameters to the observed ones.
Next, we describe how we computed these simulated X-ray series.

\subsubsection{From $\overline{CR}$ to $N$}
\label{mCRtoN}

\begin{figure}
\epsscale{1.12}
\plotone{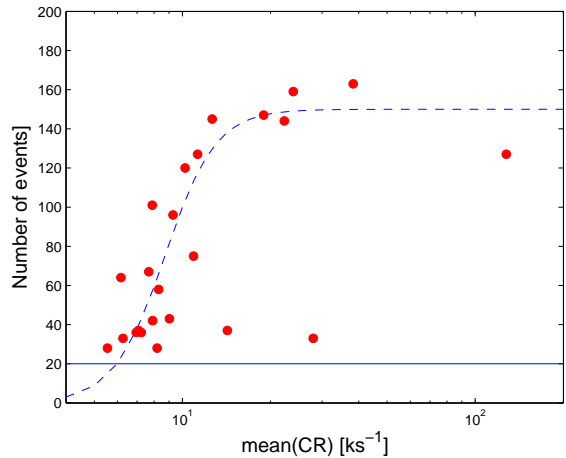}
\caption{Number of 1RXH events associated to an X-ray source as a function of
the mean net count rate ($N$ vs. $\overline{CR}$ diagram).
The solid line is the lower limit at $N$ = 20, while the dashed line is the
sigmoid function described in the main text (Eq.~\ref{eq.sigmoid}). 
\label{fig.MNmCR}}
\end{figure}

The larger the mean net count rate of an X-ray source ($\overline{CR}$) is, the
larger the number of associated events ($N$) is. 
This variation is illustrated in Fig.~\ref{fig.MNmCR}.
Except for two Mayrit stars, all the X-ray sources {\em approximately} follow
a sigmoid function defined as:
\begin{equation}
N(\overline{CR}) = \frac{A}{1 + \frac{B}{\overline{CR}^s}},
\label{eq.sigmoid}
\end{equation}
\noindent where we assume $A = 150$, $B = 5~10^4$, and $s = 5$.
Variations of $\sim$10\,\% in these parameters still provide suitable
fittings. 
The larger value of the exponent $s$ makes the sigmoid function to ``saturate''
at a maximum value of $A$ for a few tens counts per kilosecond ($N = 150$ for
$\overline{CR} \gtrsim 20$\,ks$^{-1}$), but also to trace the steep raise of $N$
at $\overline{CR} \sim 7-12$\,ks$^{-1}$.
Besides, a sigmoid function like this one gives $N \approx 0$ for null
$\overline{CR}$. 
The two outlier data points in Fig.~\ref{fig.MNmCR} correspond to Mayrit~653170
($N$ = 37, $\overline{CR}$ = 14.235\,ks$^{-1}$) and Mayrit~969077 ($N$ = 33,
$\overline{CR}$ = 27.955\,ks$^{-1}$).
Both of them displayed prominent flares among the 24 X-ray series, which
affected the determination of the mean net count rate (see below). 
The quiescent values $\overline{CR}$ correspond to the number of associated
events (and {\em vice~versa}). 
Since we are interested in generating non-variable, simulated, X-ray series
(i.e. flare-free series), we did not take into account these two outlier data
points. 
We do not know whether the relatively low number of events for the mean net
count rate in $\sigma$~Ori~AF--B is due to an observational (instrumental) bias.

\subsubsection{From $\overline{CR}$ to $CR_i$ and $\delta CR_i$}
\label{mCRtocr}

\begin{figure}
\epsscale{1.12}
\plotone{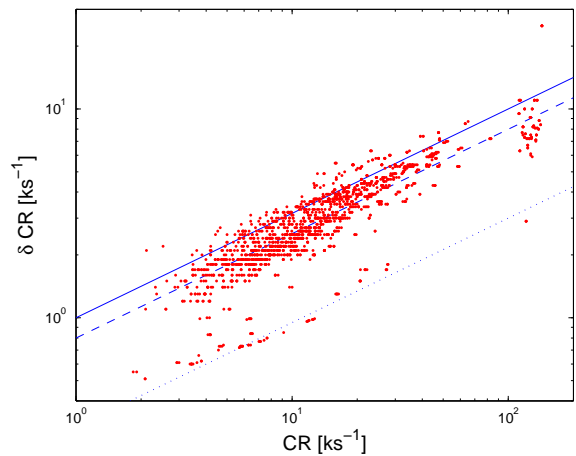}
\caption{Mean error as a function of the net count error as tabulated by the
ROSAT Team (2000) for all the 1RXH events ($\delta CR$ vs. $CR$ diagram). 
The solid, dashed, and dotted lines are for $\delta CR$ = $1.0 CR^{1/2}$,
$0.8 CR^{1/2}$, and $0.3 CR^{1/2}$, respectively.
The agglomerates of data points separated from the main data point cloud are for
$\sigma$~Ori~AF--B ($CR >$ 100\,ks$^{-1}$) and the 1992 Sep measurements
(roughly overlapping with $\delta CR$ = $0.3 CR^{1/2}$).
\label{fig.Mecrcr}}
\end{figure}

We used the MATLAB function {\tt poissrnd} to generate $N(\overline{CR})$ random
numbers, $CR_i$, following a Poissonian distribution with parameter $\lambda =
\overline{CR}$. 
There is an error, $\delta CR_i$, associated to each random value $CR_i$. 
In Fig.~\ref{fig.Mecrcr}, we plot the $\delta CR_i$ vs. $CR_i$ diagram for the
1976 investigated {\em real} events, without attending to which X-ray source
they correspond. 
The great majority of the 1995~Feb--Mar data points follow a $\delta CR \approx
0.8 CR^{1/2}$ relation.
In contrast, the 1992~Sep data points, with a better accuracy, follow a
$\delta CR_i \approx 0.3 CR_i^{1/2}$ relation.
However, they represent about 1\,\% of all considered points and their
contribution to the computation of $\chi^2$ is very small.
Thus, we assumed that the 1995~Feb--Mar relation is right for all
{\em simulated} data points, as a general rule.
Numerically, each random value $CR_i$ and its associated error satisfy the
following expression:
\begin{equation}
\delta CR_i (CR_i) = 0.8 CR_i^{1/2}.
\label{eq.critodeltacri}
\end{equation}
\noindent Each collection of pairs of values $CR_i$ and $\delta CR_i$, $i$ =
1...$N$, represent a simulated X-ray series (temporal information is 
unnecessary).
The factor 0.8 corresponds to an average observational efficiency of
$\sim$71\,\% (X-ray light curves without gaps have efficiencies of 100\,\%;
``gaps'' are observations that gave no counts for each source).

\subsubsection{From $\overline{CR}$ to $\chi^2$}
\label{mCRtochi2}

From Table~\ref{table.xrayparameters}, there are more X-ray sources with lower
mean net count rates than with higher ones.
The distribution of net count rates approximately follow a power law
$n(\overline{CR}) \propto \overline{CR}^{-0.5}$ (this is roughly the bright
end of the $\sigma$~Orionis X-ray luminosity function).
We generated 10$^5$ values of $\overline{CR}$ following this distribution and
with the restriction max($\overline{CR}$) = 200\,ks$^{-1}$.
For each value $\overline{CR}_j$, where $j$ = 1...10$^5$, we computed $N_j$
(Section~\ref{mCRtoN}) and $CR_{j,i}$ and $\delta CR_{j,i}$, where $i$ =
1...$N_j$ (Section~\ref{mCRtocr}).
Finally, with the previous values, we derived the parameters $\sigma_{CR,j}$,
$\overline{\delta CR_{j}}$, and $\chi^2_j$ (Eq.~\ref{eq.chi2}) for each
simulated series $j$.
Fig.~\ref{fig.chi2mCR} shows the subsequent $\chi^2_j$ vs. $\overline{CR}_j$
diagrams for the 10$^5$ simulated series and the 24 real series.

\subsubsection{The variability criterion}

\begin{figure*}
\epsscale{2.51}
\plottwo{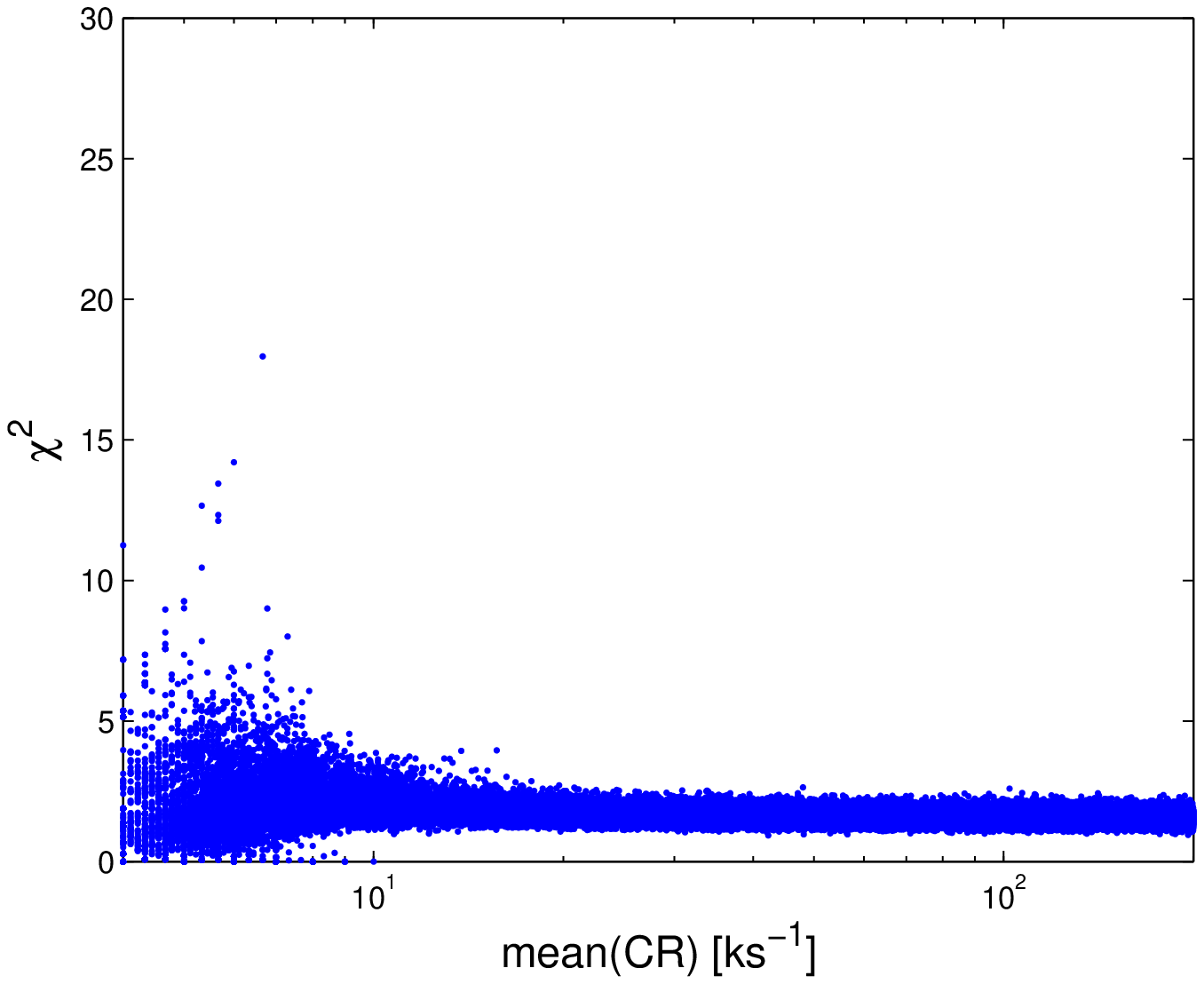}{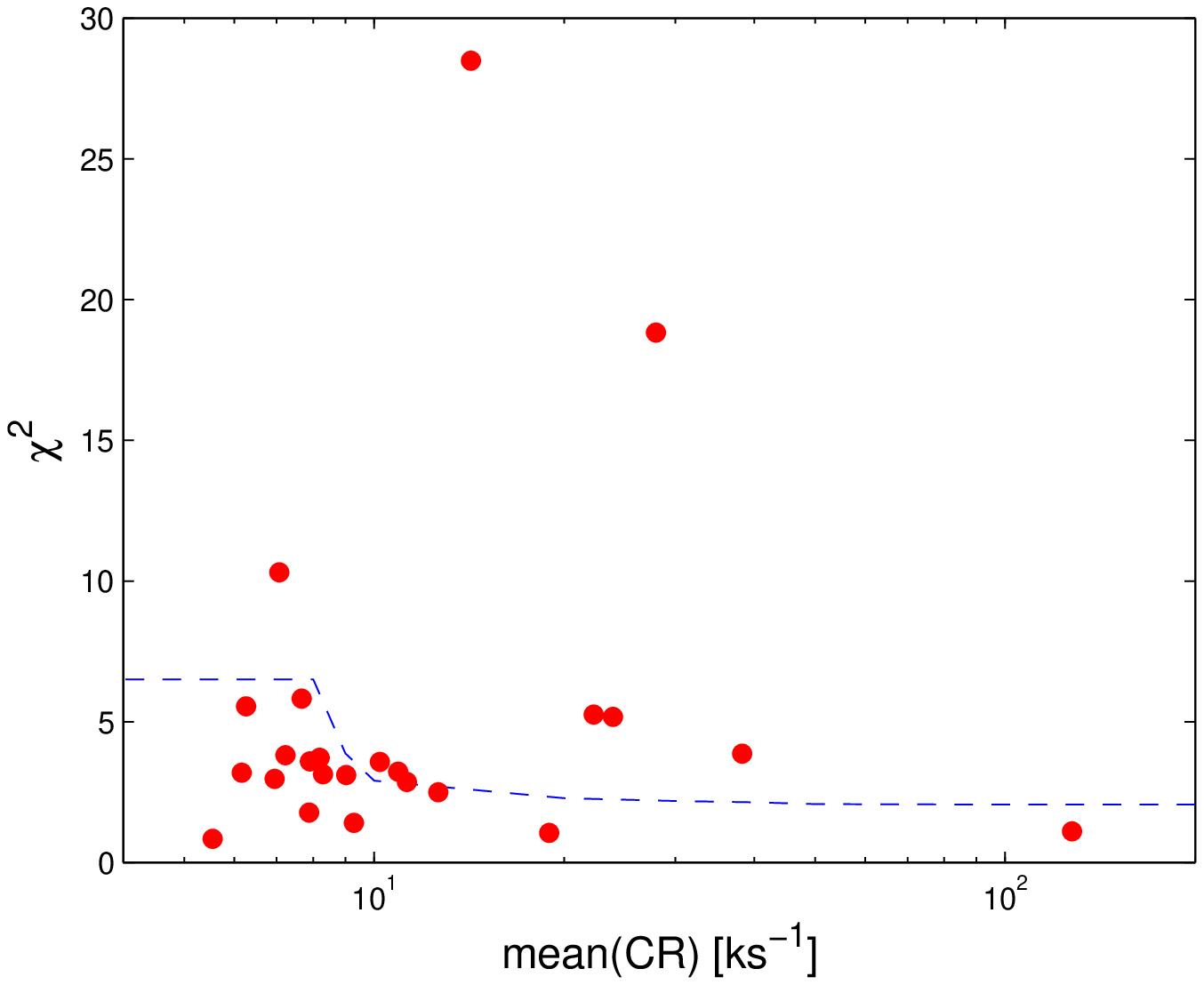}
\caption{{\em Left:} $\chi^2$ as a function of the mean net count rate
($\chi^2_j$ vs. $\overline{CR}_j$ diagram) for the 10$^5$ X-ray simulated
series.  
{\em Right:} Same as left window, but for the 24 X-ray real series.
X-ray sources above the dashed line have probabilities larger than 99\,\% of
being actual variables.
\label{fig.chi2mCR}}
\end{figure*}

We splitted the full interval of net count rates (from 4 to 200\,ks$^{-1}$) of
the simulated data into left panel in Fig.~\ref{fig.chi2mCR} in a number
of bins, where we computed the value of $\chi^2$ that leaves above it only 1\,\%
of the data points.  
In other words, every data point above this $\chi^2$ boundary has a probability
$p_{\rm var} >$ 99\,\% of being a true variable X-ray source.
From larger to smaller average net count rates, the boundary is stable at
$\chi^2 \approx$ 2.05 from $\overline{CR}$ = 200 down to 50\,ks$^{-1}$, where it
starts to increase softly up to $\chi^2 \approx$ 2.90 at $\overline{CR}$ =
10\,ks$^{-1}$. 
Then, there is a steep raise  up to an approximate value of $\chi^2$ =
6.5$\pm$0.5 at 4--8\,ks$^{-1}$ (the plateau value of the $\chi^2$ boundary in
the lowest energy interval is not as well constrained as in other intervals).
Just for clarity, the $\chi^2$ boundary is plotted only in right panel in
Fig.~\ref{fig.chi2mCR}.

\section{Results}
\label{results}

The average of the accurate data taken in 1992~Sep is consistent with the mean
quiescent level in 1995~Feb--Mar for everyone of the 24 X-ray sources (i.e.
there are no yearly variable sources from our HRI data only). 
However, there are six X-ray sources that clearly depart from the 1\,\% $\chi^2$
boundary in Fig.~\ref{fig.chi2mCR} and have, therefore, probabilities $p_{\rm
var} \gg$ 99\,\% of being variables in scales of days. 
Of the six sources, five are stars and one is the galaxy 2E~1456.
Besides, there are other four stars with $\overline{CR}$ =
10.2--11.3\,ks$^{-1}$ that fall at the variability criterion boundary (three
above, one below) and shall also be considered.
Next, we discuss the variable and non-variable sources in~detail.

\subsection{Variable sources}
\label{variable.sources}

\subsubsection{Mayrit~653170: an X-ray flaring classical T~Tauri star}
\label{M653170}

Mayrit~653170 (RU~Ori, Haro 5--15) is a long-time known classical T~Tauri star.
First discovered as a photometric variable in 1906 from Sonneberg (Kinnunen \&
Skiff 2002) and recovered by Haro \& Moreno (1953) as a strong H$\alpha$
emitter, the star went almost unnoticed for nearly one century until Wolk (1996)
found its X-ray emission. 
Mayrit~653170 has Li~{\sc i} in absorption (Caballero 2006; Sacco et~al. 2008),
He~{\sc i} $\lambda$10830\,{\AA} and Pa$\gamma$ $\lambda$10938\,{\AA} in
emission (Gatti et~al. 2008), a late-K or early-M spectral type, a radial
velocity consistent with membership in the $\sigma$~Orionis cluster
(Gonz\'alez-Hern\'andez et~al. 2008 and references above), and a spectral energy
distribution typical of disk-host stars (Hern\'andez et~al. 2007). 

The {\em ROSAT} quiescent level of Mayrit~653170 laid at $CR$ =
7$\pm$3\,ks$^{-1}$.
However, on 1995~Mar~7, the space mission detected five X-ray events with $CR$ =
59$\pm$5\,ks$^{-1}$ (i.e. eight times above the quiescent level).
The star recovered the quiescent level the following day, when three events with
$CR <$ 6\,ks$^{-1}$ were detected.

\subsubsection{Mayrit~783254 and its violent long-lived flare}
\label{M783254}

Mayrit~783254 (2E~1455, $M$ = 1.5$\pm$0.1\,$M_\odot$), although it is one of the
brightest cluster stars in the optical (Caballero 2007a) and in X-rays
(Franciosini et~al. 2006), has been poorly investigated.
Wolk (1996) determined a K0 spectral type and found both H$\alpha$ and Li~{\sc
i} in absorption.
It has no mid-infrared flux excess (Luhman et~al. 2008).
Mayrit~783254 displayed a long-lived X-ray flare that lasted at least three
days during the {\em ROSAT} observations.  
The flux increase, of a factor six, was smaller than for Mayrit~653170, but
still very clear in the X-ray light curve in top panel in Fig.~\ref{fig.X7}.
Five events at $CR$ = 113$\pm$11\,ks$^{-1}$ on 1995~Mar~11 marked the maximum of
the flare.
At this moment, Mayrit~783254 was the second brightest source in the cluster,
with only 11\,\% less soft X-ray flux than $\sigma$~Ori~AF--B.

The increasing factor and shape of the Mayrit 783254 flare were similar to those
observed in some stars of the Taurus star-forming complex (Giardino  et~al.
2006; Franciosini et~al. 2007). 
However, to the contrary of the Taurus stars, whose rise phases last less than 
one day, here we observe that the net count rate increased during a minimum of
three days.  
We suspect that the rapid decrease in the count rate after the maximum is due to
the occultation of the flaring region due to the star rotation, and that the
flare could have lasted for more than five or six days.  
Such long-duration X-ray flaring events have been observed in some stars of
Orion (e.g. Favata et~al. 2005). 
In those cases, the events were associated with long structures (loops)
anchored at the inner part of disks.  
However, a similar event has also been studied by Crespo-Chac\'on et~al. (2008)
in a star of the TW~Hydra association, 
in which the long-duration rise phase is atributed to the propagation of the
energy of the first flare along a very large arcade (e.g. Poletto et~al. 1988). 
In any case, we have no temporal and spectral resolution enough to study the
case of Mayrit~783254 in detail and we cannot distinguish between both scenarios
(a long loop vs. an~arcade).

\subsubsection{Mayrit~969077: another flare star}
\label{M969077}

Mayrit~969077 (2E~1487) first appeared in the {\em Einstein} catalogs by
Harris et~al. (1994) and Moran et~al. (1996).
There is no spectroscopic information available\footnote{Due to a typographical
error, Caballero (2008b) incorrectly tabulated Mayrit~969077 as having H$\alpha$
in emission and Li~{\sc i} in absorption.} and it also lacks a disk (Luhman
et~al. 2008). 
It felt out of the {\em XMM-Newton} observations by Franciosini et~al. (2006)
and, consequently, of the narrower {\em Chandra} observations by Skinner
et~al. (2008).
The quiescent X-ray level of Mayrit~969077 seems to be very low (at $CR$ =
10\,ks$^{-1}$ or lower), that made the star to be undetected during the first
half of the 1995~Feb--Mar run.
However, since Mar~17 to Mar~23, Mayrit~969077 produced a strong flare with a
maximum of $CR$ = 59$\pm$7\,ks$^{-1}$ on Mar~20.
The shape of the $\sim$6\,d-long flare was quite symmetric.
In this case, the lack of a disk suggests that the event might be due to an 
overposition of flares in an arcade. 
Large high-energy flares with durations of $\sim$3--6\,d, although rare, have
been detected in neighbouring stars with a ``vigorous rotationally driven
magnetic dynamo'', such as the RS~CVn-type stars UX~Ari, $\sigma$~Gem, and
V711~Tau observed with the {\em Extreme Ultraviolet Explorer} ({\em EUVE})
mission (Osten \& Brown 1999; Sanz-Forcada et~al. 2003). 
RS~CVn-type stars are close binaries ($P_{\rm orb} \sim$ 1--14\,d), so we cannot
rule out a binary hypothesis for Mayrit~969077.

\subsubsection{The bright star Mayrit~863116}

Mayrit~863116 (RX~J0539.6--0242) is a G5--K0-type, weak-lined T~Tauri star with
cosmic lithium abundance, H$\alpha$ filled in with emission, radial velocity
consistent with membership in $\sigma$~Orionis, and a broad cross-correlation
function (probably due to its high rotational velocity of $v \sin{i}$ =
150\,km\,s$^{-1}$ or to unresolved binarity; Alcal\'a et~al. 1996, 2000).  
Although it is the 19th brightest $\sigma$~Orionis star in the optical, it moves
to the ninth position in the near-infrared (Caballero 2008b).
This is due to its very red $V_T - K_{\rm s}$ color for its magnitude, of
2.25$\pm$0.05\,mag, which may be indicative of the presence of a surrounding
disk (Caballero 2007a). 
However, Hern\'andez et~al. (2007) failed to detect any flux excess with {\em
Spitzer} IRAC (4.5--8.0\,$\mu$m) and MIPS (24\,$\mu$m).

The X-ray series of Mayrit~863116 shown in top panel of Fig.~\ref{fig.X8}
displayed a flare on 1995~Mar~25 with a peak of $CR$ = 81$\pm$7\,ks$^{-1}$, at
about four times above the approximate quiescent level. 
This level seems to be variable in its turn, with short time scale variations
from 10 to 30\,ks$^{-1}$ around the approximate average value of 20\,ks$^{-1}$.

\subsubsection{The faint star Mayrit~156353}
\label{M156353}

Mayrit~156353 ([SWW2004]~36) is a photometric cluster member candidate detected
by Sherry et~al. (2004) with a moderate X-ray emission measured by Franciosini
et~al. (2006).
The last authors estimated an M1 spectral type from optical/near-infrared colors
and empirical color-magnitude relations.
They also classified Mayrit~156353 as a cluster member showing significant
variability not clearly attributable to flares. 
It is the faintest star in Table~\ref{table.opticalcounterparts} at all
wavelengths and is located relatively close to the cluster center ($\rho \sim$
2.6\,arcmin).
On 1995~Mar~8, the X-ray flux of Mayrit~156353 peaked with three events of $CR$
= 24$\pm$3\,ks$^{-1}$, about five times above the low quiescent level at $CR
\sim$ 5\,ks$^{-1}$.

\subsubsection{The galaxy 2E~1456}
\label{2e1456}

The abnormally blue $i-J$ and red $J-K_{\rm s}$ colors, strong X-ray emission,
and extended point-spread-function of the {\em Einstein} source 2E~1456 led
Caballero (2008b) to classify it as a galaxy in the background of
$\sigma$~Orionis. 
Afterwards, L\'opez-Santiago \& Caballero (2008) investigated its spectral
energy distribution from 0.3 to 7.5\,keV using {\em XMM-Newton} data (source
NX~32) and confirmed its extragalactic nature.
The net count rates of 2E~1456 in our {\em ROSAT} data varied from 20 to
60\,ks$^{-1}$, approximately, with time scales of variations of a few days. 
Unfortunately, the X-ray series of 2E~1456 obtained with {\em XMM-Newton}
by Franciosini et~al. (2006) and L\'opez-Santiago \& Caballero (2008) could not
be searched for short-scale variability due to several reasons.

The confusion of extragalactic sources with $\sigma$~Orionis member candidates
is not rare (Caballero et~al. 2008), but 2E~1456 being the second brightest
X-ray source in the area after $\sigma$~Ori~AF--B turns out remarkable, if not
singular.  
2E~1456 probably harbours an active galactic nucleus (AGN -- Rees 1984; Elvis
et~al. 1994). 
The power-law photon index derived by L\'opez-Santiago \& Caballero (2008),
$\Gamma$ = 1.82, is consistent with 2E~1456 being a Seyfert galaxy with a
supermassive black hole (e.g. Nandra \& Pounds 1994). 
Since the highest energy photons in these galaxies are believed to be created
by inverse Compton scattering by a high temperature corona near the compact
black hole, the observed X-ray variability in time scales of a few days is not
difficult to explain. 

\subsubsection{Four possible variable stars with $p_{\rm var} \sim$ 99\,\%}
\label{fourpossiblevariables}

From Fig.~\ref{fig.chi2mCR}, there are four stars with $CR$ =
10.2--11.3\,ks$^{-1}$ and relatively large $\chi^2$ values corresponding to
probabilities of variability at the 1\,\% $\chi^2$ boundary.
Two of them, both with $p_{\rm var} \gtrsim$ 99\,\%, are chemically
peculiar early-type stars: Mayrit~42062 ($\sigma$~Ori~E; B2Vp) and Mayrit~306125
(HD~37525; B5Vp). 
Their variable X-ray emission would be difficult to explain if they were single
(Skinner et~al. 2008), especially in the case of the flare of Mayrit~42062.
Previously, strong X-ray flares in Mayrit~42062 had been reported by Groote \&
Schmitt (2004; using these very same HRI observations) and Sanz-Forcada
et~al. (2004). 
Such flares are typical in young late-type stars, but virtually missing in
early-type stars like Mayrit~42062.
Besides, Skinner et~al. (2008) detected no large flares on it, although
they saw a possible sinusoidal variation with a period consistent with the
stellar rotation period $P \sim$ 1.19\,d.
After decades of unfruitful searches (Landstreet \& Borra 1978; Groote \& Hunger
1982; Drake et~al. 1994; Townsend et~al. 2005 and references therein), Bouy
et~al. (2009) finally discovered the low-mass companion at only $\rho
\approx$ 0.330\,arcsec to Mayrit~42062.
This late-type companion is likely the origin of the detected flares (it is
also plausible that the B2Vp star itself is the only source of the X-ray
emission, including the variability -- ud-Doula et~al. 2006; Townsend et~al.
2007).  
Likewise, Mayrit~306125 is another binary: Caballero (2006) resolved a companion
of the primary about 0.5\,mag fainter at $\rho$ = 0.47$\pm$0.04\,arcsec (see
also Caballero 2005).
However, from the magnitude difference in this case, the companion
Mayrit~306125~B must be a late B- or an early A-type star.
The combined X-ray lightcurve of the binary system does not show apparent flare
events, but a soft variation from 5 to 17\,ks$^{-1}$ in time scales of tens of
days. 
Variable colliding wind shocks of the two components may play an important
r\^ole in the X-ray emission of the system.

The other two possible variable stars are Mayrit~203039 ([W96]~4771--1049;
$p_{\rm var} \gtrsim$ 99\,\%) and Mayrit~114305 ([W96]~4771--1147~AB;
$p_{\rm var} \lesssim$ 99\,\%).
Skinner et~al. (2008) also measured variability probabilities larger than
99 and 97\,\%, respectively, from {\em Chandra} data.
On the one hand, Mayrit~203039 has a K5 spectral type, Li~{\sc i} in
absorption, and H$\alpha$ in chromospheric emission (Wolk 1996). 
A short-duration flare, that persisted for no less than the
last 4\,h of their observations, was discovered by Franciosini et~al. (2006).
The time series from our data in middle panel of Fig.~\ref{fig.X4} does not
evidence apparent flare events in Mayrit~203039, but quasi-periodic variations
from 5 to 22\,ks$^{-1}$ in time scales of tens of days.
Lower-amplitude, higher-frequency variations could be superimposed to the main
trend.
On the other hand, Mayrit~114305 displays quite similar properties, except for
an earlier spectral type (K0), a spectroscopic binary status (Wolk 1996), and a
possible evolved disk (Hern\'andez et~al. 2007).
Given its proximity to the cluster center ($\rho$ = 1.9\,arcmin), it has been
investigated in other X-ray surveys in $\sigma$~Orionis (Sanz-Forcada et~al.
2004; Franciosini et~al. 2006; Caballero 2007b). 
The time series of Mayrit~114305, shown in top panel of Fig.~\ref{fig.X3},
diplayed two clear flares on 1995~Mar~9 and~19 with peaks of $CR$ = 26$\pm$5
and 25$\pm$4\,ks$^{-1}$, respectively (five X-ray events each).

\subsection{Non-variable stars}

In our data, there are two stars with large mean net count rates and low values
of $\chi^2$ and are, therefore, suitable examples of non-variable X-ray stars.
They are $\sigma$~Ori~AF--B and Mayrit~789281 (2E~1454; but see
Section~\ref{discussion}).
The absence of mid-term variability in the massive triple system confirms
previous results by Groote \& Schmitt (1994; see also Skinner et~al. 2008 for a
detailed discussion on the origin of its X-ray emission).
However, Mayrit~789281 is a late-G or early-K T~Tauri star with cosmic lithium
abundance and H$\alpha$ in broad emission (Caballero 2006;
Gonz\'alez-Hern\'andez et~al. 2008).
Its radial velocity matches the average one of the ``Group~1'' of young stars
that contaminate the $\sigma$~Orionis cluster and belong to an overlapping
not-so-young stellar population in the Orion Belt (Jeffries et~al. 2006). 
The X-ray time series of Mayrit~789281 maintained very stable at $CR$ =
19$\pm$4\,ks$^{-1}$, with no outlier data points.

The remaining 12 stars are below the boundary of $p_{\rm var}$ = 99\,\% and are
classified in this work as non-variable stars with our conservative
classification criterion.
Nevertheless, the 12 of them have $\overline{CR} <$ 10\,ks$^{-1}$, where the
boundary of $p_{\rm var}$ = 99\,\% was not well constrained.
Furthermore, the X-ray time series of a few stars, like Mayrit~105249
([W96]~rJ053838--0236) and Mayrit~344337 ([W96]~4771--1097), display flare-like
events and have values of $\chi^2$ close to the boundary of $p_{\rm var}$ =
99\,\% (the two of them were classified as cluster members showing
significant variability not clearly attributable to flares by Franciosini et~al.
2006; Mayrit~105249 did show flaring activity in Skinner et~al. 2008 data).
Instead of being suspicious of the actual variability of the faint flare star
Mayrit~156353 (also with $\overline{CR} <$ 10\,ks$^{-1}$;
Section~\ref{M156353}), we are suspicious of the non-variability of some of
our 12 ``non-variable'' stars.
In any case, we prefer being cautious and claiming the variability and possible
variability from our data of only the other nine stars (and one galaxy) in
Section~\ref{variable.sources}.

\section{Discussion}
\label{discussion}

Using an {\em XMM-Newton} dataset, Sanz-Forcada et~al. (2004) and
Franciosini et~al. (2006) reported strong flare events in 12 $\sigma$~Orionis
stars. 
Ten of them were too faint for our HRI {\em ROSAT}; 
the other two bright X-ray sources were Mayrit~203039 and Mayrit~42062
(classified in our work as possible variables;
Section~\ref{fourpossiblevariables}).  
Another five stars in Franciosini et~al. (2006) showed significant variability
not clearly attributable to flares.
Of them, four have been investigated here and only one, Mayrit~207358, had no
hint of variability from our $\chi^2$ analysis.
Besides, of the nine X-ray HRI stars in common with Skinner et~al. (2008), four
had variability probabilities larger than 97\,\% in their work.
Three of them are among the four possible variable stars with $p_{\rm var} \sim$
99\,\% (Section~\ref{fourpossiblevariables}) and the remaining one is
Mayrit~105249, which we classified as a questionable non-variable star.
All these facts inidicate that our ``possible variable stars'' likely vary. 

The variability criterion that we used in Section~\ref{results}, although
it was computed through an apparently complicated method, gives the same
results as if we had searched for flares in the light curves with the naked eye. 
We also performed single sample goodness-of-fit hypothesis
Kolmogorov-Smirnov tests on the HRI time series by assuming hypothetical
Poissonian cumulative distributions functions of mean $\lambda = \overline{CR}$
(we used the MATLAB functions {\tt poisscdf} and {\tt kstest}).
In last column in Table~\ref{table.xrayparameters}, we provide the corresponding
Kolmogorov-Smirnov $p$-values (a source with a $p$-value $p_{\rm
K-S}(\overline{\lambda})$ has a probability $1 - p_{\rm
K-S}(\overline{\lambda})$ of being variable).
All except five sources do not reject the null hypothesis at significance level
$\alpha \equiv p_{\rm K-S}(\overline{\lambda})$ = 10$^{-7}$.
The five of them are among the six variable sources in
Section~\ref{variable.sources}. 
The sixth variable there, the faint star Mayrit~156353, has a $p$-value less
than 2\,10$^{-4}$.
Another six stars have $p$-values below the 5\,10$^{-3}$ threshold (with K-S
variability probabilities larger than 99.5\,\%).
All of them except one are discussed here as probable variables or questionable
non-variables.
The only inconsistency between our $\chi^2$ and Kolmogorov-Smirnov analyses
is the apparently stable young star Mayrit~789281, which has $p_{\rm
K-S}(\overline{\lambda}) \approx$ 1.3\,10$^{-4}$.
Another variability criteria, such as maximum likelihood block algorithm,
would have led us to comparable outcomes if they were to be performed
(Wolk et~al. 2005; Stelzer et~al. 2007; Albacete Colombo et~al. 2007b).  

The existence of nine HRI variable stars [five if not accounting for the
possible variables in Section~\ref{fourpossiblevariables}] among a list of 23
stars leads to a frequency of mid-term X-ray variability of 39\,\% [22\,\% if
not accounting for the possible variables]\footnote{Because of the two different
ways of computing the frequency of X-ray variability, we do not give error bars.
If computed, it should be as in Burgasser et~al.~(2003).}.
Except in two cases (the possible variable stars Mayrit~306125 and
Mayrit~203039), the origin of the X-ray variability is the presence of flaring
events. 

Five [seven] of the six [ten] sources with $p_{\rm var} \gg$ 99\,\% [$p_{\rm
var} \gtrsim$ 99\,\%] (including the galaxy) are among the seven [eleven]
brightest X-ray sources. 
This is an evidence of a bias towards the non-detection of variability among the
faintest sources. 
Therefore, the actual frequency of mid-term X-ray variability in
$\sigma$~Orionis must be larger than 39\,\% [22\,\%]. 
Observational selection and statistical effects in the determination of X-ray
variability have been discussed in detail by, e.g., Stelzer et~al (2000; with
{\em ROSAT}) and Albacete Colombo et~al. (2007a; with {\em Chandra}).

In the literature, there have been numerous studies of X-ray variability and
flaring activity of stars with {\em ROSAT}:
M- and F7--K2-type dwarfs in the solar neighbourhood (Marino et~al. 2000, 2002),
stars in clusters significantly older than $\sigma$~Orionis (such as
$\alpha$~Persei, $\tau \sim$ 90\,Ma; Prosser et~al. 1996), or stars in the
Taurus-Auriga-Perseus sky region (with Taurus star forming region, Pleiades, and
Hyades-like ages; Stelzer et~al. 2000).
With the arrival of {\em XMM-Newton} and {\em Chandra}, larger samples of stars
in young clusters are observed (Preibisch \& Zinnecker 2002 in IC~348;
Gagn\'e et~al. 2004 in $\rho$~Ophiuchi; Marino et~al. 2005 in IC~2391).
The most comprehensive searches for X-ray variability have been carried out in
the Orion Nebula Cluster with {\em Chandra} by Feigelson et~al. (2002), Favata
et~al. (2005), and Wolk et~al. (2005).
Reported frequencies of X-ray variabilities range between $<$5\,\% in the
field to $\sim$46\,\% in IC~2391 ($\tau \sim$ 40\,Ma).
Our measured X-ray variability of 39\,\% [22\,\%] lays within this range.
However, due to the different spectral response of the X-ray missions, stellar
ages, and depth, length, and temporal resolution of observations, our derived
frequency of X-ray variability is not {\em directly} comparable to these
determinations. 
We insist that the data and analysis presented in this paper are sensitive
to relatively high amplitude variability on timescales in excess of a day, but
insensitive to other types of variability.
Nevertheless, our work complements satisfactorily previous X-ray variability
studies in young stars in the $\sigma$~Orionis cluster, especially those by
Franciosini et~al. (2006) and Skinner et~al. (2008).
Although our HRI data prevented us from detecting low amplitude variability and
in scales shorter than a few hours, {\em ROSAT} archival data are to date the
only way to investigate X-ray variations in scales of days. 

Likewise, we could not perform any study of the flare energy distribution due to
the obvious limitations of the HRI {\em ROSAT} observation in terms of time
coverage (one observation of about 2.2\,ks $\approx$ 0.025\,d every day),
spectral coverage (0.1--2.4\,keV), and spectral resolution (there is no defined
bands in the 1RXH catalogue).  
Results on this topic in other star forming regions investigated with {\em
Chandra} and {\em XMM-Newton} have been presented by other authors, such as
Albacete Colombo et~al. (2007b) in the the Cyg~OB2 association and Orion Nebula
Cluster and Stelzer et~al. (2007) in~Taurus.  

Taking into account the compiled spectrophotometric information and the
$iJHK_{\rm s}$ magnitudes in the Mayrit catalog (Caballero 2008b), we looked 
for particular features of the X-ray variable stars, especially the presence of
(accretion) circumstellar disks.   
There is a tendency of disk-host classical T~Tauri stars to be less X-ray
luminous than disk-free young stars (Neuh\"uaser et~al. 1995 in Taurus;
Franciosini et~al. 2006 and Caballero 2007b in $\sigma$~Orionis). 
More importantly, disk-free stars, which rotate faster than stars magnetically
locked by disks, are more X-ray active (G\"udel 2004; Preibisch et~al. 2005).  
The enhancement of the activity by fast stellar rotation is also applied to
$\sigma$~Orionis.  
While the frequency of disks in the cluster at almost all mass intervals lies at
about 50\,\% (Caballero 2007a and references therein; Sacco et~al. 2008; Luhman
et~al. 2008), only one (the classical T~Tauri star Mayrit~653170 in
Section~\ref{M653170}) of the nine [five] stars with $p_{\rm var}
\gtrsim$ 99\,\% [$p_{\rm var} \gg$ 99\,\%] is a class~II object with
mid-infrared flux excess due to a disk.
In other words, the frequency of mid-term X-ray variable stars with disks in our
sample is 11\,\% [20\,\%].
If both disk-free and disk-host $\sigma$~Orionis stars had the same X-ray
properties (i.e. the same rotational velocity and subsequent activity),
we would expect four or five [two or three] ($\sim$50\,\%) class~II
objects with mid-term variability instead of only one.
Apart from the disk-host star Mayrit~653170, there is no trend between X-ray
variability and optical/near-infrared colors or H$\alpha$ emission. 
A detailed study of the frequency of flares and X-ray properties in stars with
and without disk in a large sample of members of the $\sigma$~Orionis cluster
observed with HRC-I onboard {\em Chandra} is being done by the authors and
will be presented in a forthcoming paper.

\section{Summary}

Using Virtual Observatory tools and HRI {\em ROSAT} public data, we have
investigated the X-ray variability in scale of days of 23 stars in the
young $\sigma$~Orionis cluster. 
Catalogued data covered more than 30\,d in 1995~Feb--Mar.
The time series of five stars displayed clear flare events (with probabilities
$p_{\rm var} \gg$ 99\,\% of being variable); 
several of these flares were violent and lasted for up to six days.
Another four stars seemed to be also variable in scales of days.
All the flaring activity identifications are new except for Mayrit~203039
(a weak-lined T~Tauri star) and Mayrit~42062 ($\sigma$~Ori~E and its close faint
companion), for which we confirm the display of flares. 
Since our data are insensitive to low-amplitude variations or with time
scales shorter than one day, the actual frequency of X-ray variables among
$\sigma$~Orionis is larger than 39\,\%.
Classical T~Tauri stars with disks are less X-ray variable (or, alternatively,
less X-ray luminous) than disk-free young stars, which presumably rotate faster.
Besides, we have identified an AGN-host galaxy with an intense, variable X-ray
emission.

\acknowledgments

We are grateful to the anonymous referee, J.~F. Albacete Colombo, and J.
Sanz-Forcada for helpful discussion and advice. 
Financial support was provided by the Universidad Complutense de Madrid,
the Comunidad Aut\'onoma de Madrid, the Spanish Ministerio Educaci\'on y
Ciencia, and the European Social Fund under grants:
AyA2005-02750,				
AyA2005-04286, AyA2005-24102-E,		
AyA2008-06423-C03-03, 			
AyA2008-00695,				
PRICIT S-0505/ESP-0237,			
and CSD2006-0070. 			
This research has made use of the SIMBAD, operated at Centre de Donn\'ees
astronomiques de Strasbourg, France, and the NASA's Astrophysics Data System.

\clearpage

\begin{deluxetable}{l cc cc c ccc c l c}
\tablecolumns{9}
\tablewidth{0pc}
\tablecaption{Bright {\em ROSAT} sources in $\sigma$~Orionis: 
our X-ray parameters
\label{table.xrayparameters}}
\tablehead{
\colhead{Name}		& \colhead{$\overline \alpha$}	& \colhead{$\overline \delta$}    & \colhead{$\sigma_\alpha$}	& \colhead{$\sigma_\delta$}	& \colhead{$N$} & \colhead{$\overline{CR}$}	& \colhead{$\sigma_{CR}$}	& \colhead{$\overline{\delta CR}$}	& \colhead{$\chi^2$} & Var.	& $p_{\rm K-S}$	\\ 
			& \colhead{(J2000)}	& \colhead{(J2000)}     & \colhead{[arcsec]}		& \colhead{[arcsec]}		&		& \colhead{[ks$^{-1}$]}  	& \colhead{[ks$^{-1}$]}  	& \colhead{[ks$^{-1}$]}  		& 		& 		& $(\overline{\lambda})$ 	}
\startdata
\object{Mayrit~AB} 	& 084.686413  		& --02.599601        	& 1.4460       			& 1.6507          		& 127       	& 127.61       			& 8.2144       			& 8.4512       				& 1.0972 	& No 		& 0.30		\\
\object{Mayrit~42062} 	& 084.696596     	& --02.594071       	& 1.7106       			& 2.0294           		& 75       	& 10.908       			& 5.3903       			& 2.4067       				& 3.2179 	& Yes? 		& 1.5~10$^{-4}$	\\
\object{Mayrit~105249} 	& 084.658954      	& --02.610452       	& 2.6008       			& 2.2433           		& 67       	& 7.6745       			& 5.4678       			& 1.9472       				& 5.8211 	& No: 		& 1.7~10$^{-5}$	\\
\object{Mayrit~114305} 	& 084.659967      	& --02.581494       	& 2.3823       			& 2.8311          		& 145       	& 12.627       			& 4.5136       			& 2.5663       				& 2.4952 	& No? 		& 1.4~10$^{-4}$	\\
\object{Mayrit~156353} 	& 084.681297      	& --02.556549       	& 1.4855       			& 2.3738           		& 37       	& 7.0708       			& 5.4232       			& 1.8357       				& 10.309 	& Yes 		& 1.7~10$^{-4}$	\\
\object{Mayrit~157155} 	& 084.704907      	& --02.639707       	& 1.9022       			& 1.9348           		& 28       	& 5.5457       			& 1.3996         		& 1.7400      				& 0.8330 	& No 		& 0.64		\\
\object{Mayrit~180277} 	& 084.636298     	& --02.593567       	& 2.8343       			& 2.9797           		& 64       	& 6.1658       			& 2.7547       			& 1.7486         			& 3.1900 	& No 		& 0.063		\\
\object{Mayrit~203039} 	& 084.722404      	& --02.555867       	& 2.3429        		& 2.4370          		& 127       	& 11.259       			& 4.2434       			& 2.4978       				& 2.8527 	& Yes? 		& 0.0011	\\
\object{Mayrit~207358} 	& 084.684171   		& --02.542337   	& 2.2816       			& 3.7694          		& 101       	& 7.8854       			& 2.4518        		& 2.0580        			& 1.7690 	& No 		& 0.12		\\
\object{Mayrit~260182} 	& 084.683889      	& --02.672494       	& 2.2552       			& 2.1461           		& 36       	& 7.2283       			& 5.4087        		& 1.9650       				& 3.8026 	& No 		& 8.8~10$^{-5}$	\\
\object{Mayrit~285331} 	& 084.647531      	& --02.530589       	& 2.2976        		& 1.7840           		& 36       	& 6.9514       			& 3.4081          		& 1.9000       				& 2.9707 	& No 		& 0.29		\\
\object{Mayrit~306125} 	& 084.756245      	& --02.649166       	& 2.0980       			& 3.2594          		& 120       	& 10.207       			& 3.4418       			& 2.3645       				& 3.5731 	& Yes? 		& 2.0~10$^{-4}$	\\
\object{Mayrit~344337} 	& 084.649288      	& --02.511315        	& 2.1160        		& 2.7610           		& 33       	& 6.2621       			& 3.8114       			& 1.7788       				& 5.5412 	& No: 		& 0.0044	\\
\object{Mayrit~528005} 	& 084.700105      	& --02.452731        	& 2.5920       			& 2.3517           		& 58       	& 8.2898       			& 2.9667       			& 2.3005       				& 3.1352 	& No 		& 0.024		\\
\object{Mayrit~615296} 	& 084.531778     	& --02.524775       	& 2.9341       			& 2.6624           		& 96       	& 9.2787       			& 2.3336       			& 2.5375       				& 1.3998 	& No 		& 0.077		\\
\object{Mayrit~634052} 	& 084.825398     	& --02.490467       	& 3.5232       			& 4.0022           		& 42       	& 7.9033       			& 3.1513       			& 2.4155       				& 3.5873 	& No 		& 0.12		\\
\object{Mayrit~653170} 	& 084.716724      	& --02.779516       	& 3.5521       			& 2.1357           		& 37       	& 14.235       			& 18.075       			& 2.4011       				& 28.487 	& Yes 		& 8.9~10$^{-18}$\\
\object{Mayrit~750107} 	& 084.886696		& --02.662064	        & 3.7316       			& 3.8277           		& 28       	& 8.1939        		& 3.4150       			& 2.5711       				& 3.7112 	& No 		& 0.018		\\
\object{Mayrit~783254} 	& 084.475637		& --02.657996	        & 3.8509			& 2.7251			& 159		& 23.902			& 17.473			& 4.2384				& 5.1667 	& Yes 		& 8.5~10$^{-19}$\\
\object{Mayrit~789281} 	& 084.469905      	& --02.559010       	& 3.9973       			& 3.8648          		& 147       	& 18.939       			& 3.9355       			& 3.8728       				& 1.0496 	& No 		& 1.3~10$^{-4}$	\\
\object{Mayrit~822170} 	& 084.725095      	& --02.825619       	& 3.4521        		& 3.7930           		& 43       	& 9.0223       			& 3.8657       			& 2.6023       				& 3.1078 	& No 		& 0.032		\\
\object{Mayrit~863116} 	& 084.902644       	& --02.704998       	& 3.3539       			& 3.6062          		& 144       	& 22.266       			& 12.898       			& 4.0868       				& 5.2504 	& Yes 		& 1.2~10$^{-12}$\\
\object{Mayrit~969077} 	& 084.949897        	& --02.539979       	& 4.6934       			& 4.8872           		& 33       	& 27.955       			& 16.276       			& 4.6606       				& 18.825 	& Yes 		& 9.1~10$^{-8}$	\\
\object{2E~1456} 	& 084.483910      	& --02.753679       	& 2.7119       			& 2.9404          		& 163       	& 38.294       			& 9.5044       			& 5.4239       				& 3.8583 	& Yes 		& 1.6~10$^{-9}$	\\
\enddata
\end{deluxetable}

\begin{deluxetable}{ll cc c l}
\tablecolumns{9}
\tablewidth{0pc}
\tablecaption{Bright {\em ROSAT} sources in $\sigma$~Orionis: 
Optical counterparts
\label{table.opticalcounterparts}}
\tablehead{
\colhead{Name}	& \colhead{Alternative}	& \colhead{$\alpha$}	& \colhead{$\delta$}    & \colhead{$J$}		& \colhead{Remarks}			\\ 
		& \colhead{name}	& \colhead{(J2000)}     & \colhead{(J2000)}     & \colhead{[mag]}	& 					}
\startdata
Mayrit~AB     & $\sigma$~Ori~A+B+F+IRS1	& 084.686518		& --02.600047	        &  4.020$\pm$0.010	& OB, mIR				\\ 
Mayrit~42062 	& $\sigma$~Ori~E (a+b)	& 084.696659		& --02.594594	        &  6.974$\pm$0.026	& OB					\\ 
Mayrit~105249 	& [W96]~rJ053838--0236	& 084.659273		& --02.610675	        & 11.158$\pm$0.026	& Li~{\sc i}, H$\alpha$ 		\\ 
Mayrit~114305 	& [W96]~4771--1147~AB	& 084.660359		& --02.581950	        &  9.097$\pm$0.027	& Li~{\sc i}, H$\alpha$			\\ 
Mayrit~156353 	& [SWW2004]~36		& 084.681473		& --02.557045	        & 11.716$\pm$0.028	& XX ([FPS2006] 76)			\\ 
Mayrit~157155 	& [W96]~rJ053849--0238	& 084.704888		& --02.639512	        & 11.389$\pm$0.023	& Li~{\sc i}, H$\alpha$ 		\\ 
Mayrit~180277 	& [W96]~rJ053832--0235b	& 084.636844		& --02.594223	        & 11.544$\pm$0.027	& XXX ([FPS2006] 49)			\\ 
Mayrit~203039 	& [W96]~4771--1049	& 084.722390		& --02.556384	        & 10.607$\pm$0.026	& Li~{\sc i}, H$\alpha$ 		\\ 
Mayrit~207358 	& [W96]~4771--1055	& 084.684345		& --02.542674	        & 10.877$\pm$0.027	& Li~{\sc i} 				\\ 
Mayrit~260182 	& [W96]~4771--1051	& 084.684299		& --02.672143	        & 11.363$\pm$0.026	& Li~{\sc i}, H$\alpha$, class~II	\\ 
Mayrit~285331 	& [W96]~rJ053835--0231	& 084.647775		& --02.531016	        & 11.360$\pm$0.026	& XX ([FPS2006] 60)			\\ 
Mayrit~306125 	& HD~37525~AB		& 084.756235		& --02.649011	        &  8.131$\pm$0.030	& OB					\\ 
Mayrit~344337 	& [W96]~4771--1097	& 084.649467		& --02.512033	        & 11.245$\pm$0.026	& Li~{\sc i}, H$\alpha$			\\ 
Mayrit~528005 	& [W96]~4771--899~AB	& 084.700149		& --02.453943	        & 10.156$\pm$0.023	& Li~{\sc i}, H$\alpha$, class~II	\\ 
Mayrit~615296 	& 2E~1459	  	& 084.532698		& --02.525389	        & 10.566$\pm$0.027	& Li~{\sc i}, H$\alpha$			\\ 
Mayrit~634052 	& [W96]~4771--0598	& 084.825298		& --02.491239	        & 10.721$\pm$0.027	& XX ([FPS2006] 156) 			\\ 
Mayrit~653170 	& RU~Ori	  	& 084.716695		& --02.778798	        & 11.518$\pm$0.026	& Li~{\sc i}, H$\alpha$, class~II	\\ 
Mayrit~750107 	& [W96]~rJ053932--0239	& 084.885707		& --02.662225	        & 10.820$\pm$0.024	& Li~{\sc i}				\\ 
Mayrit~783254 	& 2E~1455	  	& 084.476687		& --02.658291	        &  9.255$\pm$0.020	& XXX ([FPS2006] 3)			\\ 
Mayrit~789281 	& 2E~1454	  	& 084.470982		& --02.559558	        &  9.991$\pm$0.027	& Li~{\sc i}, H$\alpha$			\\ 
Mayrit~822170 	& RX J0538.9--0249	& 084.725444		& --02.824937	        & 10.829$\pm$0.026	& Li~{\sc i}, H$\alpha$			\\ 
Mayrit~863116 	& RX~J0539.6--0242  	& 084.902263		& --02.704768	        &  8.462$\pm$0.027	& Li~{\sc i}, H$\alpha$			\\ 
Mayrit~969077 	& 2E~1487  		& 084.949336		& --02.540242	        & 10.969$\pm$0.026	& Li~{\sc i}, H$\alpha$			\\ 
2E~1456 	& \nodata	  	& 084.484608		& --02.753637	        & 15.398$\pm$0.073	& Galaxy				\\ 
\enddata
\end{deluxetable}

\clearpage

\clearpage

\clearpage

\end{document}